\documentclass{article}
\usepackage[utf8]{inputenc}
\usepackage{url,hyperref,lineno,microtype,subcaption}
\usepackage{setspace}
\usepackage{multirow}
\usepackage{enumerate}
\usepackage{verbatim}
\usepackage{array}
\usepackage{xcolor}
\usepackage[numbers]{natbib}
\usepackage{graphicx}
\graphicspath{ {./images/} }
\usepackage{amsmath}
\usepackage{cleveref}
\usepackage{multicol}
\usepackage{amsmath,blkarray,booktabs,bigstrut}
\usepackage{enumitem}
\usepackage{multirow}
\usepackage{booktabs}
\usepackage{tabularx}
\usepackage{adjustbox}
\usepackage{rotating}
\usepackage[font=footnotesize, labelfont=bf]{caption}
\usepackage{subcaption}
\usepackage{attachfile}
\usepackage{authblk}
\usepackage{geometry}
\usepackage{float}
\usepackage{subcaption}
\usepackage{soul}

\DeclareRobustCommand{\Niloofar}[1]{ {\begingroup\sethlcolor{green}\hl{[Niloofar] :  #1}\endgroup} }

\DeclareMathOperator*{\argmax}{arg\,max}

\hypersetup{colorlinks,linkcolor={black},citecolor={black},urlcolor={red}}

\title{SGC: A semi-supervised pipeline for gene clustering using self-training approach in gene co-expression networks}
\author[]{Niloofar Aghaieabiane\thanks{corresponding author, niloofar.aghaieabiane@njit.edu}}
\author[]{Ioannis Koutis\thanks{ikoutis@njit.edu}}
\affil[]{{\small Department of Computer Science, New Jersey Institute of Technology, Newark, NJ, United States}}
\date{}

\begin{document}

\begin{abstract}
    A widely used approach for extracting information from gene expression data employ the construction of a {\em gene co-expression network} and the subsequent application of algorithms that discover network structure. In particular, a common goal is the computational discovery of gene clusters, commonly called {\em modules}. When applied on a novel gene expression dataset, the quality of the computed modules can be evaluated automatically, using {\em Gene Ontology enrichment}, a method that measures the frequencies of Gene Ontology terms in the computed modules and evaluates their statistical likelihood.  %Existing widely used pipelines rely on relatively simple statistical and mathematical tools that do not reflect recent algorithmic and mathematical advances. 
    In this work we propose \textsc{SGC} a novel pipeline for gene clustering based on relatively recent seminal work in the mathematics of spectral network theory. \textsc{SGC} consists of multiple novel steps that enable the computation of highly enriched modules in an unsupervised manner. But unlike all existing frameworks, it further incorporates a novel step that leverages Gene Ontology information in a {\em semi-supervised} clustering method that further improves the quality of the computed modules. Comparing with already well-known existing frameworks, we show that SGC results in higher enrichment in real data. In particular, in \( 12 \) real gene expression datasets, SGC outperforms in all except one.
\end{abstract}

%\begin{abstract}
%Gene co-expression network is a systematic framework which is widely used to extract biological knowledge. The goal is to cluster the genes such that those with similar expression pattern fall within the same cluster commonly call {\em module}. Gene ontology enrichment is a technique that is used for module evaluation. The higher the enrichment, the better the module. Existing pipelines rely on relatively simple statistical and mathematical tools to find the modules. More importantly, modules are found based on internal information of clusters by hierarchical clustering. Here, we propose a novel pipeline self-training gene clustering (SGC) for gene co-expression network analysis. SGC incorporates multiple steps and find the final module based on the information of kmeans clustering, gene ontology, and spectral graph theory knowledge. The highlight is it changes the problem from unsupervised to semi-supervised. Comparing with already well-known existing frameworks, we show that SGC results in more module enrichment in real data.
%\end{abstract}

\maketitle
%%%%%%%%%%%%%%%%%%%%%%%%%%%%%%%%%%%%%%%%%%%%%%%%%%%%%%%%%%%%%%%%%%%%%%%%%%%%%%%%%%%%%%%%%%%%%%%%%%%%%%%%%%%%%%%%%%

\section{Introduction}
\label{sec:introduction}

\par High throughput gene expression data enables gene functionality understanding in fully systematic frameworks. Gene module detection in Gene Co-expression Networks (GCNs) is a prominent such framework that has generated multiple insights, from 
unraveling the biological process of plant organisms~\cite{Emamjomeh2017} and essential genes in microalgae~\cite{Panahi2021}, to
assigning unknown genes to biological functions~\cite{Ma2018} and recognizing disease mechanisms~\cite{Parsana2019}, e.g.~for coronary artery disease~\cite{Liu2016}.

\par GCNs are graph-based models where nodes correspond to genes and the strength of the link between each pair of nodes is a measure of similarity in the expression behavior of the two genes~\cite{Tieri2019}. The goal is to group the genes in a way that those with similar expression pattern fall within the same network cluster, commonly called {\em module}~\cite{Gat2003, Sipko2017}. GCNs are constructed by applying a similarity measure on the expression measurements of gene pairs. Genes are then clustered using unsupervised graph clustering algorithms. Finally the modules are analyzed and interpreted for gene functionality~\cite{niloo2021}.

\par The de facto standard automatic technique for module quality analysis is Gene Ontology (GO) enrichment, a method that reveals if a module of co-expressed genes is enriched for genes that belong to known pathways or functions. Enrichment is a measure of module quality and the module-enriching GO terms can be used to discover biological meaning~\cite{Khatri2005, niloo2021, Botia2017, Russo2018}. Statistically, in a given module, this method determines the significance of the GO terms for a test query by associating {\em p-values }. The query includes the test direction, either ``underrepresented" (under) or ``overrepresented" (over), and three ontologies; ``biological process" (BP), ``cellular component" (CC), and ``molecular function" (MF). {\em p-values} are derived based on the number of observed genes in a specific query with the number of genes that might appear in the same query if a selection performed from the same pool were completely random. In effect, this values identify if the GO terms that appear more frequently than would be expected by chance~\cite{Khatri2005}. As usual the smaller the {\em p-value} the more significant the GO term.

%\Yiannis{The following paragraph talks about``de facto" GO enrichment, when the above paragraph also did  that. What is the role of this paragraph then? It feels like the author has not read the paragraph above, and just puts random stuff in the text - so I added one sentence, above and we can delete the one below because it makes little sense, even after the correction. Keep in mind that the reader may not know about GO enrichment, so we need to tell them very quickly what it is and how it works, instead of just saying platitudes like ``unravel the associations''. We have to be very clear about what we say, and everything we say needs to have an intended purpose.}.
%\par Gene ontology (GO) enrichment is the de facto standard technique for module analysis and quality~\cite{Khatri2005, niloo2021}. It can determine which biological functions and genes are strongly associated with a module~\cite{Rhee2008}. The task in GCNs is to unravel of the associations between genes in a module and inquiry ontology~\cite{Sebastian2014} and scores the modules on the basis of their GO enrichment. The more significant the GO term enrichment, the higher the score for the module.

\subsection{Background on existing GCN frameworks}
\label{sec:Background}

\par Several frameworks and algorithms have been developed for GCNs construction and analysis such as~\cite{Zhang2005general, Lan2008, petal2015,coseq,CoXpress, Botia2017, Russo2018}. Among them Weighted Correlation Network Analysis (WGCNA)~\cite{ Lan2008}, is still the most widely accepted and used framework for module detection in GCNs~\cite{niloo2021, Botia2017, Russo2018, Liu2016, Hou2021}. WGCNA 
uses the Pearson correlation of gene expressions to form a `provisional' network and then powers the strength values on its links so that the network conforms with a ``scale-freeness'' criterion. The final network is constructed by adding to the provisional network additional second-order neighborhood information, in the form of what is called  topological overlap measure (TOM). Finally, WGCNA uses a standard hierarchical clustering (HC) algorithm to produce modules~\cite{dynamicTreeCut}. 

\par In recent years, there has been a growing interest to enhance WGCNA and multiple frameworks have been proposed as a modification of this framework. These pipelines mainly utilize an additional step in the form of either pre-processing or post-processing to WGCNA. Co-Expression Modules identification Tool (CEMiTool) is a pipeline that incorporates an extra pre-processing step to filter the genes using the inverse gamma distribution~\cite{Russo2018}. In another study, it is shown that a calibration pre-processing step in raw gene expression data results in increased GO enrichment~\cite{niloo2021}. Two other frameworks, the popular CoExpNets~\cite{Botia2017} and K-Module~\cite{Hou2021}, have utilized k-means clustering~\cite{KMeans} as a post-processing step to the output of WGCNA. Finally, in a comparative study, it is revealed that CEMiTool has advantage over WGCNA~\cite{Cheng2020}. 

%\Yiannis{Does that comparative study include CoExpNets? What is the reason we bring it up here?} \Niloofar{No it does not examine CoExpNets}
    
\subsection{Our framework: Self-trained Gene Clustering}
\label{sec:ourFramework}

\par We have developed Self-training Gene Clustering (SGC), a user-friendly R package for GCNs construction and analysis. Its integration with Bioconductor makes it easy to use and apply. SGC differentiates itself from WGCNA and other pipelines in three key ways: (i) It constructs a network without relying on the scale-freeness criterion which has been controversial~\cite{SChaefer2017,Raya2006, Lima2009,Broido2019,Clote2020}. 
(ii) It clusters the network using a variant of spectral graph clustering that has been proposed relatively recently in seminal work~\cite{Cheeger}. (iii) It incorporates `self-training', a supervised clustering step that GO enrichment information obtained from the previous step and further enhances the quality of the modules. To our knowledge SGC is the first pipeline that uses GO enrichment information as supervision. 

%~\cite{Liu2021}.
%\Yiannis{It is not clear what this reference does. Throwing-in a claim like 'finer modules' that has not been defined or mean anything to the reader is something we should avoid. We need to think about every sentence we write, and put it tightly in its context, where the reference means something and plays a role.}

\subsubsection{SGC: The workflow}
\label{sec:SGCoverview}

\par
\begin{figure}[h]
\includegraphics[width=1\textwidth]{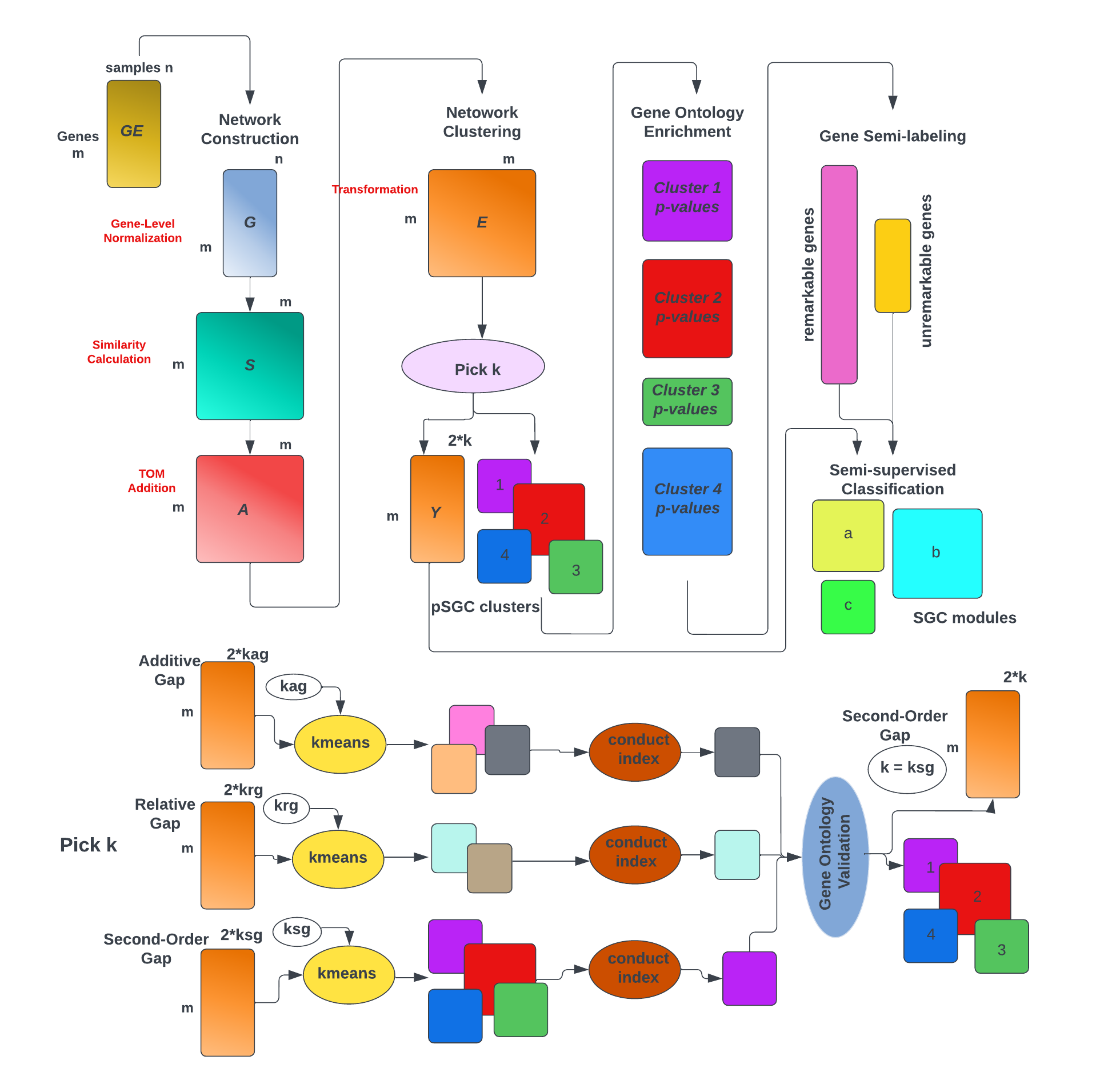}
\caption{The SGC pipeline for gene clustering in gene co-expression networks. SGC takes the gene expression matrix \( GE \) and outputs clusters and their refinements to modules after a semi-supervised classification steps. The steps for picking the number of clusters \( k \) is drawn below the main pipeline. \label{fig:flowchart}}
\end{figure}

 The workflow of SGC is illustrated in figure~\autoref{fig:flowchart}; in what follows we give an overview of SGC and also point to the corresponding sections containing more details. SGC takes as input a gene expression matrix \( GE \) with \( m \) genes and \( n \) sample and performs the following five main steps:

\smallskip

\noindent {\em \textbf{Network Construction}}: Each gene vector, i.e.~each row in matrix \( GE \) is normalized to a unit vector; this results to a matrix \( G \). Next, the Gaussian kernel function is used as the similarity metric to calculate \( S \) in which \( 0 \leq s_{i,j} = s_{j,i} \leq 1\) and \( s_{i,j} \)~shows the similarity value between gene \( i \) and \( j \). Then, the second-order neighborhood information will be added to the network in the form of topological overlap measure~(TOM)~\cite{Zhang2005general}. The result of this step is an $m\times m$ symmetric adjacency matrix \( A \). 
[Section~\ref{subsubsec:NetworkConstruction}]

\smallskip

\noindent {\em \textbf{Network Clustering}}: Matrix \( A \) is used to define and solve an appropriate eigenvalue problem. The eigenvalues are used to determine three potential values $(k_{ag}, k_{rg},k_{sg})$ for the number of clusters $k$ ([Section~\ref{subsubsec:networkClustering}]). For each such value of $k$, SGC computes a clustering of the network, by applying the kmeans algorithm on an embedding matrix $Y$ generated from $2k$ eigenvectors. In each clustering it finds a \textit{test cluster}, defined as the cluster with the smallest conductance index. The three test clusters are evaluated for GO enrichment, and SGC picks the clustering that yielded the test cluster with highest GO enrichment. This clustering is the output of the Network Clustering step, and its clusters are the {\em initial clusters}. [Section~\ref{subsubsec:networkClustering}]

%The number of clusters $k$ is automatically determined, and $2k$ eigenvectors are used to compute the matrix $E$ which is a geometric embedding for the $m$ genes in $2k$ dimensions. [\ref{subsubsec:networkClustering}]
%will be transformed into the new embedding~\( E \) and then \( 2 * k \) dimensions of the it will be selected for the next steps where \( k \) is the number of clusters, call it \( Y \). 
%To find the number of clusters $k$, SGC computes three potential values of $k$ and for each such value it computes a 

%SGC employs three methods; \textit{gap}, \textit{first}, and \textit{second}. For each method, it performs kmeans to find the clusters and it then pick the cluster that has the minimum conductance index. In Ontology Cross Validation, GO enrichment is carried out for those clusters and finally SGC picks the \( k \) according to the method that its overall significant GO terms is the best. The clusters of the corresponding method is considered as the output of Network Clustering step and are called {\em initial clusters}.

\smallskip

\noindent {\em \textbf{Gene Ontology Enrichment}}: GO enrichment analysis is carried out on the initial clusters individually. [Section~\ref{sec:geneOntology}] 

\smallskip

\noindent {\em \textbf{Gene semi-labeling}}: Genes are categorized into \textit{remarkable genes} and \textit{unremarkable genes} using information derived from the GO enrichment step. For each cluster, remarkable genes are those that have contributed to GO terms that are more significant relative to a baseline. Remarkable genes are labeled according to their corresponding cluster label. Not all clusters contain remarkable genes, and thus a new number $k' \leq k$ of clusters is determined, and accordingly, $k'$ labels are assigned to the remarkable genes and to the corresponding geometric points in the embedding matrix $Y$ computed in the Network Clustering step. This defines a supervised classification problem. 

\smallskip

\noindent {\em \textbf{Semi-supervised classification}}: The supervised classification problem is solved with an appropriately selected and configured machine learning algorithm (either k-nearest neighbors~\cite{bishop}, or one-vs-rest logistic regression~\cite{bishop}) with the remarkable genes as the training set. The supervised classification algorithm assigns labels to the unremarkable genes. At the end of the this step, all the genes are fully labeled, and the final clusters called \textbf{\textit{modules}} are produced. SGC returns two set of modules, those obtained by the unsupervised Network Clustering step, and those produced by the Semi-supervised classification step. For clarity, in this study, the former and the latter are called \textbf{\textit{clusters}} and \textbf{\textit{modules}} and we denote the corresponding methods with pSGC (prior to semi-supervised classification) and SGC respectively. [Section~\ref{sec:SemiSupervisedClassification}].

\smallskip
\textbf{Remark:} Computing the Gene Ontology Enrichment is a computationally time-expensive task. The process for selecting $k$, described in the Network Clustering step, is meant to reduce the amount of computation for the GO enrichment. However, whenever the amount and time of computation is not of concern, multiple other values of $k$ can be evaluated (whenever possible independently, by parallely running computing processes). This has the potential to produced even better modules. Indeed, in the single case when our method does not outperform the baselines (see Section~\ref{sec:ResultDiscussion}), a different choice of $k$ does produce a `winning' output for our framework.

\subsubsection{A comparison of SGC with existing frameworks}
\par SGC deviates from commonly used existing pipelines for GCNs in three key ways: \\ (i) \textit{Network Construction:} While existing pipelines employ a procedure that relies on a controversial scale-freeness criterion, SGC employs a Gaussian kernel whose computation relies on simple statistics of the dataset that are not related to scale-freeness considerations. To the extent that SGC is effective in practice reveals that scale-freeness is \textbf{not} fundamental in GCNs, affirming the findings of multiple other works on biological networks. \\
(ii) \textit{Unsupervised Clustering:} Most existing pipelines employ hierarchical clustering algorithms as the main tool for unsupervised learning step. SGC first computes a spectral embedding of the GCN and then applies kmeans clustering on it. Crucially, the embedding algorithm is based on a recent breakthrough in the understanding of spectral embeddings of networks.  \\
(iii) \textit{GO-based supervised improvement:}
Existing frameworks do not make any use of GO information, except for providing it in the output.
This includes methods that work on improving the quality of a first set of `raw' clusters. SGC is the first framework that explicitly uses GO information to define a semi-supervised problem which in turn is used to find more enriched modules.

%there are controversial bottlenecks: (i) Pipelines for GCNs are highly depends on Pearson correlation as the main similarity metric along with scale-freeness criteria in which it is assumed that biological networks follow this assumption. In SGC, we digress from these steps and combine them into single step by using the Gaussian Kernel geometric metric. Gaussian kernel is a non-linear function for Euclidean distance, in which Euclidean coefficients are used as the power to raise the exponential multiplied with some normalization factor. In other words, in SGC the Gaussian kernel step corresponds the two Pearson correlation and scale-freeness steps combine. This is more appropriate for GCNs context since it prevents producing negative numbers, as the expression levels are positive values. Whereas, in PC networks it is not clear to interpret the negative values. Additionally, as we will show later, it turns out that the scale-freeness criteria is not always a valid assumption. (ii) Most of pipelines rely on hierarchical clustering algorithm as the main tool for unsupervised learning step. SGC, however, is highly depends on the kmeans clustering on the embedded data. We found out this leads to more enriched modules. (iii) None of the already existing pipeline uses the information of the GO inside their frameworks. Whereas, SGC uses this information to find more enriched modules, and converts the unsupervised problem into semi-supervised.

\section{Results and Discussion}
\label{sec:ResultDiscussion}

We present experiments that demonstrate that SGC outperforms three competing state-of-the art methods on a wide variety of datasets. 

%each {\em prominent} module contains GO terms of higher statistical significance 

\medskip 

\textbf{Experimental Setting.} We compare pSGC (i.e.~SGC without semi-supervised cluster improvement) and SGC with three pipelines (WGCNA, CoExpNets, CEMiTool) on 12 gene expression datasets: \( 4 \) DNA-microarray datasets~\cite{GSE33779, GSE44903, GSE28435, GSE38705} and \( 8 \)  RNA-sequencing datasets~\cite{GSE181225, GSE54456, GSE57148, GSE60571, GSE107559, GSE104687, GSE150961, GSE115828}. These include expression arrays with a wide range of samples from \( 5 \) to \( 511 \), various organisms, along with different units~\cite{Abbas-Aghababazadeh2018}.\footnote{ Expression units provide a digital measure of the abundance of gene or transcripts.} The datasets were downloaded from the NCBI Gene Expression Omnibus (GEO) database~\cite{GEO}. Details on the datasets are available in \autoref{tab:GSESummary}. We note that raw DNA-microarray datasets are normalized using robust multiarray analysis (RMA)~\cite{RMA} which is the most popular preprocessing step for Affymetrix~\cite{affymetrix} expression arrays data~\cite{McCall2010}. 

$\href{run:SupplementaryFile3_PipelineModuleDetail.xlsx}{Supplementary File3} $ provides a summary of the GO enrichment results for all pipelines and \( 12 \) datasets.

\begin{table}
\centering
\caption{Benchmark datasets and summary statistics of applying pipelines WGCNA, CoExpNets, CEMiTool, PSGC, SGC on them. The first \( 6 \) columns contain dataset information. Possible dataset types are: DNA-microarray (DNA) or RNA-seq (RNA). Datasets come from the the following organisms: Homo sapiens (Hs), Drosophila melanogaster (Dm),  Rattus norvegicus (Rn), Mus musculus (Mm). Units are Relative Log Expression (RLE), Reads Per Kilobase of transcript per Million mapped reads (RPM), Fragments Per Kilobase of exon per Million mapped fragments (FKPM), Trimmed Mean of M-values (TMM). The ``sft" column indicates softpower used by tested benchmarks to enforce the network to be scale-free.
For each of the 5 pipelines, k is the number of clusters and \#GO Terms indicated the number of gene ontology terms found in all modules collectively by the pipeline; in particular, a single \#GO term will appear once for each cluster where its presence exceeds a threshold of significance in term of its p-value. In the case of SGC, ``mth" denotes the method ultimately used for selecting \( k \), ag: additive gap, rg: relative gap, and sg: second-order gap. \%UNR Genes shows the percentage of the entire genes at are unremarkable. \%CH Label shows the percentage of the unremarkable genes that their labels have changed after semi-labeling step.}
\label{tab:GSESummary}

\begin{adjustbox}{width=\textwidth}

\begin{tabular}{||l c c c c c || c c || c c || c c || c c || c c c || c c||}
 \toprule
 \multicolumn{6}{|c|}{} & \multicolumn{2}{|c|}{WGCNA} & \multicolumn{2}{|c|}{CoExpNets} & \multicolumn{2}{|c|}{CEMiTool} & \multicolumn{2}{|c|}{pSGC} & \multicolumn{5}{|c|}{SGC} \\
 \cmidrule(l){7-19}
 Data & Type & Organism & \#Samples & Unit & sft &  k & \#GO Terms &  k & \#GO Terms &  k & \#GO Terms  &  k & \#GO Terms &  k & \#GO Terms & mth &  \% UNR Genes & \% CH Label \\ 

\hline
\hline
GSE181225~\cite{GSE181225} &  RNA & Hs & 5 & RLE & 26 & 48 & 7462 & 75 & 9027 & 32 & 6252 & 2 & 2598 & 2 & 2598 & ag,rg & 1\% & 0\%  \\
\hline
GSE33779~\cite{GSE33779} & DNA & Dm & 90 & probes & 14 & 22 & 5631 & 19 & 6213  & 17 & 5299 & 10 & 4144 & 7 & 3821 & ag & 56\% & 47.1\% \\
\hline
GSE44903~\cite{GSE44903} & DNA & Rn & 142 & probes & 30 & 18 & 3298 & 27 & 4705 & 14 & 3303 & 4 & 987 & 4 & 1059 & rg & 29\% & 5\%   \\
\hline
GSE54456~\cite{GSE54456} & RNA & Hs & 174 & RPKM & 30 & 31 & 9386 & 46 & 14473 & 22 & 11056 & 3 & 6004 & 3 & 600 &  ag,rg & 1\% & 1\%  \\
\hline
GSE57148~\cite{GSE57148} & RNA & Hs & 189 & FPKM & 14 & 45 & 13296 & 36 & 14110 & 33 & 12027 & 9 & 2833 & 5 & 2383 & sg  & 46\% & 24\%  \\
\hline
GSE60571~\cite{GSE60571} & RNA & Dm & 235 & FPKM & 9 & 21 & 7107 & 19 & 8622 & 16 & 5564 & 2 & 2969 & 2 & 2971 &  ag,rg & 21\% & 2\%   \\
\hline
GSE107559~\cite{GSE107559} & RNA & Hs & 270 & FPKM & 3 & 26 & 10915 & 20 & 12499  & 80 & 15952 & 14 & 5257  & 12 & 4913 & sg  & 6\% & 48.4\%  \\
\hline
GSE28435~\cite{GSE28435} & DNA & Rn & 335 & probes & 22 & 51 & 7331 & 47 & 7769 & 31 & 6566 & 2 & 2052 &  2 & 2053 &  sg & 12\% & 0\%   \\
\hline
GSE104687~\cite{GSE104687} & RNA & Hs & 377 & FPKM & 18 & 31 & 10339 & 28 & 1193 & 23 & 11369 & 2 & 6426 & 2 & 6426 &  ag,rg & 0\% & 0\%   \\
\hline
GSE150961~\cite{GSE150961} & RNA & Hs & 418 & TMM & 5 & 9 & 3619 & 18 & 606 & 17 & 4856 & 2 & 2111 & 2 & 2111 &  ag,rg & 9\% & 0\% \\
\hline
GSE115828~\cite{GSE115828} & RNA & Hs & 453 & CPM & 12 & 51 & 12611 & 10 & 7693 & 10 & 5231 & 3 & 1934 &  3 & 1926 & sg  & 33\% & 0\%  \\
\hline
GSE38705~\cite{GSE38705} & DNA & Mm & 511 & probes & 16 & 8 & 3320  & 12 & 4123 & 7 & 3308 & 6 & 2824 & 4 & 2610 & sg & 39\% & 62\%  \\

\hline
\hline

\bottomrule

\end{tabular}

\end{adjustbox}
\end{table}

\medskip

We look at the following metrics of quality. 

\par
\begin{figure}[!htb]
\centering
\begin{subfigure}{.48\textwidth}
  \centering
  \includegraphics[width=0.95\linewidth]{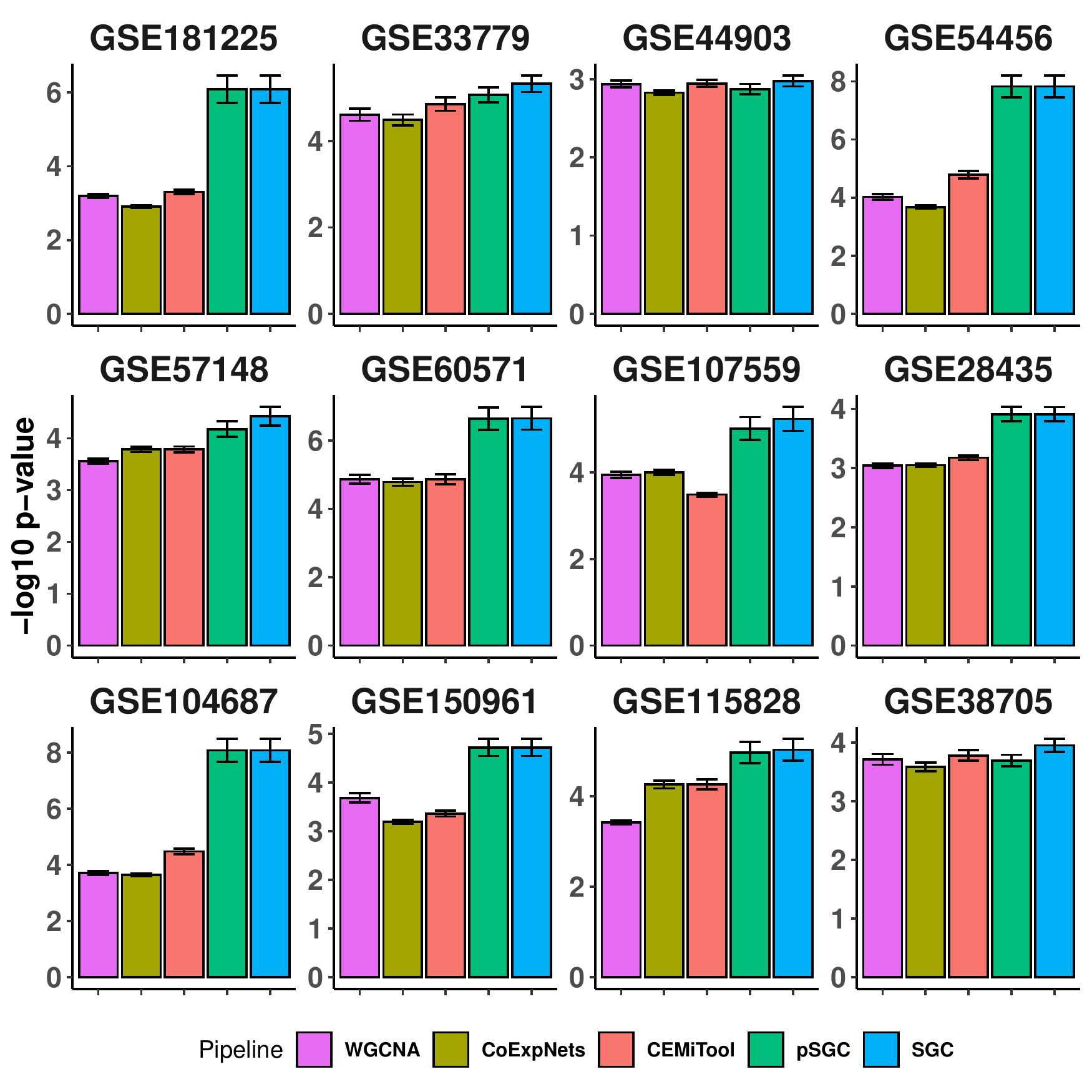}
  \caption{}
  \label{fig:barPlot}
\end{subfigure}
\begin{subfigure}{.48\textwidth}
  \centering
  \includegraphics[width=0.95\linewidth]{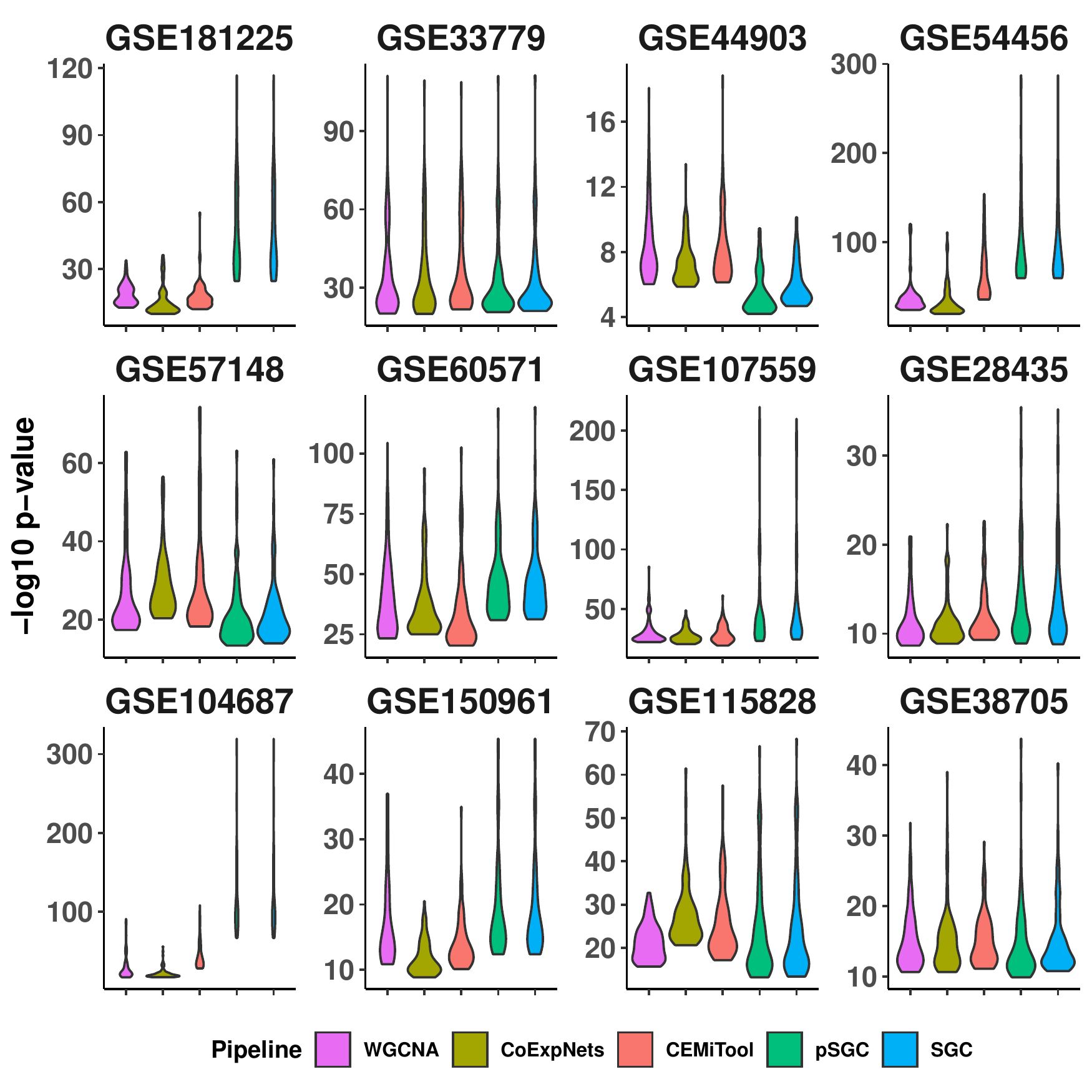}
  \caption{}
  \label{fig:violinPlot}
\end{subfigure}
\begin{subfigure}{.48\textwidth}
  \centering
  \includegraphics[width=0.95\linewidth]{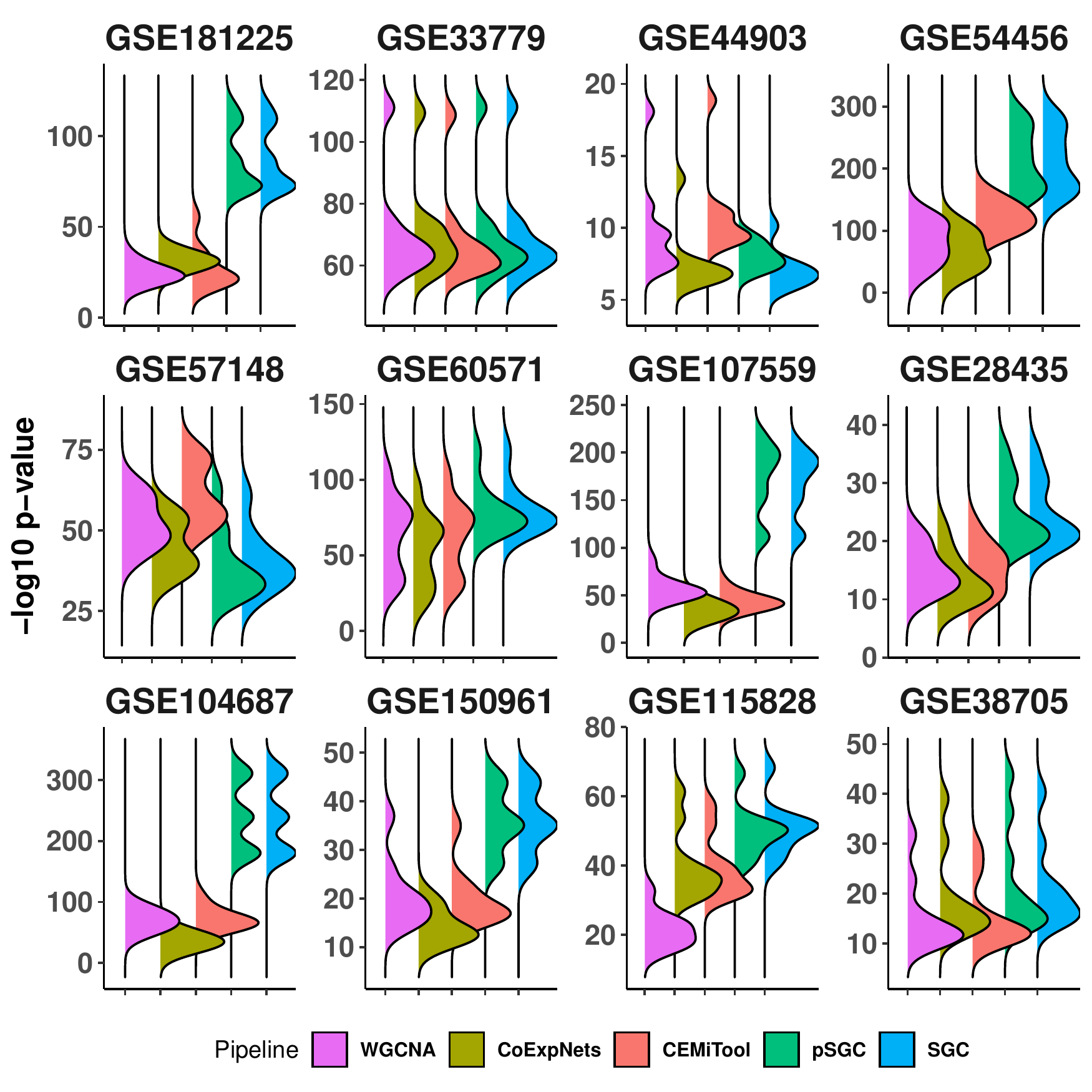}
  \caption{}
  \label{fig:densityPlot}
\end{subfigure}
\begin{subfigure}{.48\textwidth}
  \centering
  \includegraphics[width=0.95\linewidth]{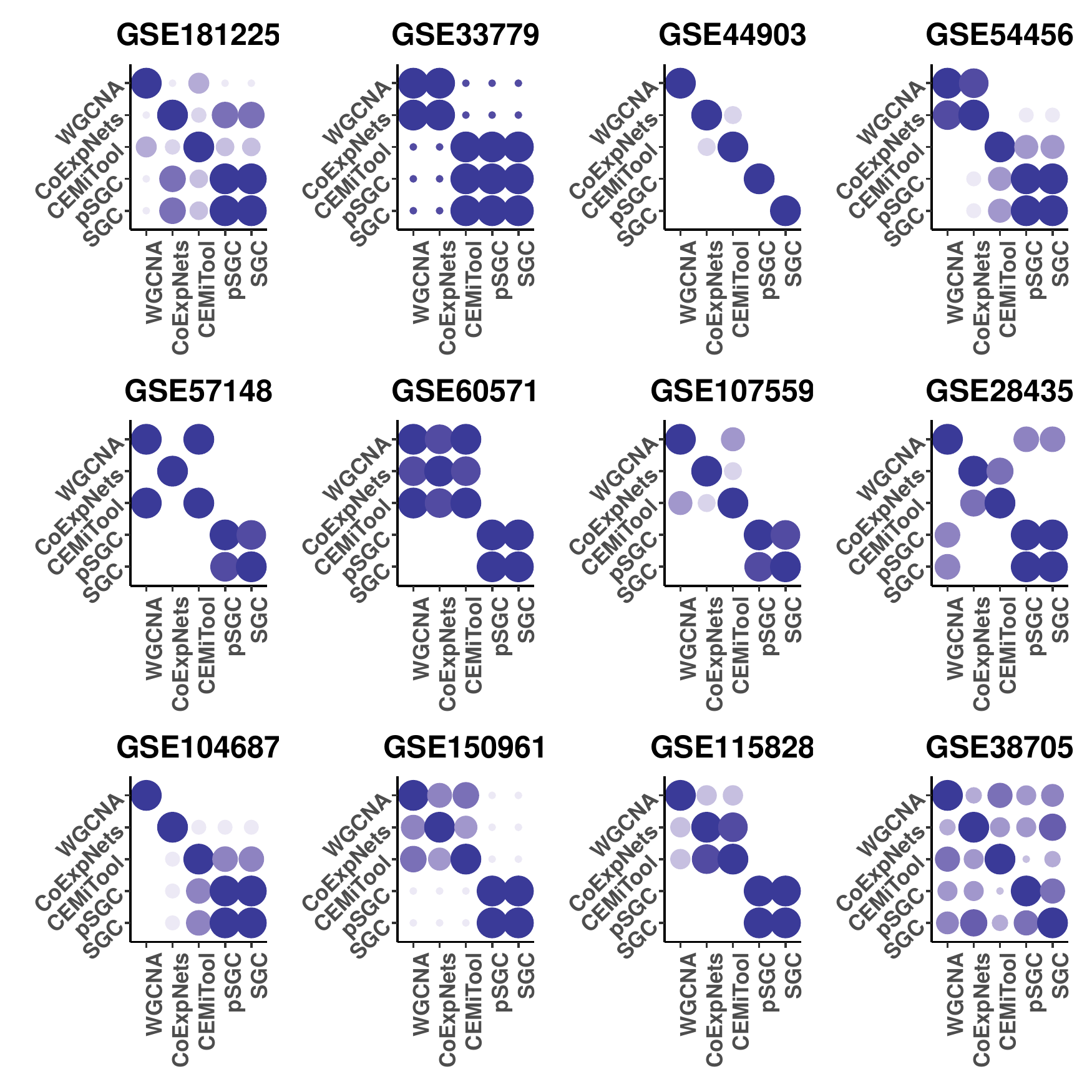}
  \caption{}
  \label{fig:OddsRatioPlot}
\end{subfigure}
\caption{Gene ontology enrichment analysis comparing WGCNA, CoExpNets, CEMiTool, pSGC, and SGC in 12 real datasets. {\em p-values} are log-transformed. The order of the pipelines from left to right are WGCNA (purple), CoExpNets(yellow), CEMiTool (orange), pSGC(green), and SGC (blue). (a) All {\em p-values} from all modules are pooled, averaged, and shown as barplot. Error bars indicated the 95\% confidence intervals that have been calculated based on the standard deviation of the {\em p-values}. (b) Top \( 100 \) most significant {\em p-values} from all modules are shown as violin plot. (c) Top \( 10 \) most significant {\em p-values} for the prominent module for each pipeline. 
%(d) Odds ratio percentage for pipeline on the data individually. The odds ratio values are discretized into 4 bins; \( 0 \) , \( \le 1 \), \( \geq 1 \), \( +Inf \). Stacks show the percentage of GO terms that fall in each bin.\label{fig:performance}
(d) Overlaps in top-100 GO terms reported by the five different frameworks. For pipeline $p$ in the $x$-axis and pipeline $q$ in the $y$-axis, position $(p,q)$ shows the number of GO terms reported by both $p$ and $q$,  among their top unique $100$ GO terms. The bigger and darker the circle, the higher the overlap.}
\end{figure}

 \textbf{i. Average Cluster Quality.} We follow the previous convention and methodology~\cite{Song2012, CCor2016}, and evaluate performance by comparing the {\em p-values} returned by pipelines. Let \( p_{i,j} \) be the \( i \)th order {\em p-value} calculated for module \( j \). Then, the quality of module \( j \) is defined as \( q_j = -(\sum_{j=1}\log_{10} p_{i,j})/n_j \) where \( n_j \) is the number of GO terms found in module \( j \). Finally, the quality of framework \( f \) is defined as \( Q_f= (\sum_{i=1}^{k}q_i )/k \) where \( k \) is the number of modules in \( f \). The results are shown in~\autoref{fig:barPlot}. SGC outperforms the three baselines on all datasets, and the same is true for pSGC, with the exception of GSE38705. We can also see that SGC is at least as good as pSGC on all datasets, and in 6/12 of the datasets it improves the module quality. 

%We found out, in all \( 12 \) gene expression data, pSGC followed by SGC perform better than three tested methods~\autoref{fig:performance}. More precisely, let \( p_{i,j} \) be the \( i \)th order {\em p-value} calculated for module \( j \). Then, the quality of module \( j \) is defines as \( q_j = -(\sum_{j=1}\log_{10} p_i,j)/n_j \) where \( n_j \) is the number of GO terms found in module \( j \). Finally, the quality of framework \( f \) is defines as \( Q_f= (\sum_{i=1}^{k}q_i )/k \) where \( k \) is the number of modules in \( f \).~\autoref{fig:barPlot} shows the pipeline performance across the benchmark dataset. The higher the bar, the finer the pipeline performance. As it seen, in all the data module quality for pSGC followed by SGC outperforms the other pipeline except GSE44903. When the majority of the genes are {\em remarkable} in a data, pSGC and SGC have similar performance. Otherwise, SGC improves the module quality. Note that {\em p-values} are log-transformed. 

\par \textbf{ii. Most significant GO terms.} The summary evaluation includes all p-values for the GO terms, as reported by GoStats, but here we focus on the top \( 100 \) {\em p-values} for each pipeline. ~\autoref{fig:violinPlot} reports these p-values in the form of `violin' plots. The y-axis indicates the significance of each GO term in terms of p-value. The top GO terms in pSGC and SGC have a higher p-value than the corresponding top terms of the other frameworks except for datasets GSE44903 and GSE57148; in GSE57148 only CEMiTool does better than SGC. It can be also observed that in \( 5 \) datasets (GSE181225, GSE54456, GSE107559, GSE28435, GSE104687), SGC is {\em dominant} to other frameworks, as the {\em least significant} GO term found by SGC is more significant than the majority of GO terms founds by the other frameworks. In datasets (GSE150961, GSE11582, GSE60571, and GSE38705) the `violin' for pSGC and SGC tends to be higher relative to the other frameworks. In two datasets (GSE33779, GSE38705) the three pipelines have similar performance.

\par \textbf{iii. GO terms of most significant module.} We consider as most significant or {\em prominent}, the module that contains the GO term containing the highest p-value. We then consider the 10 most significant GO terms in the prominent module and we show their p-values in~\autoref{fig:densityPlot}. We observe that, even when restricted to the prominent module,  pSGC and SGC report more significant terms than other methods, on all datasets except GSE44903 and GSE57148; in GSE57148 only CEMiTool is better than SGC. In 6 of the datasets, pSGC and SGC are astonishingly better than the other frameworks. Figure~\ref{fig:OddsRatioPlot} in fact shows that the most significant GO terms reported by SGC are mostly different from those reported by the baseline frameworks.

%Then, we dive into prominent modules. We say, module \( m \) in pipeline \( p \) is {\em prominent} if the most significant {\em p-value} in \( m \) is more significant than other all {\em p-values} in other modules of \( p \). In next effort, the distribution of the \( 10 \) top GO terms in prominent module of each framework is illustrated for the all GE data individually.~\autoref{fig:densityPlot} shows the result. As usual, the higher the density the more significant the {\em p-values}. The GO terms found in prominent module of pSGC and SGC is higher than the three other methods except in GSE44903 and GSE57148. Similar to~\autoref{fig:violinPlot}, in GSE57148, CEMiTool only performs better than pSGC and SGC. Similar to~\autoref{fig:violinPlot} in \( 6 \) dataset, pSGC and SGC are dominant to others astonishingly. In two data, the pipeline almost have similar enrichment, and in rest of the data, they still perform better than the three tested framework. This result suggests that pSGC clusters and SGC modules are more enriched in compare with WGCNA, CoExpNets, and CEMiTool. Specially we found out for the top GO terms, that make differences in enrichment, modules of pSGC and SGC are noticeably more significant in compare with other three tested methods. In summary, we found that SGC followed by pSGC outperform the three tested methods across all benchmark datasets with only one exception. In effect, modules of the SGC are more enriched than pSGC clusters, and the clusters of pSGC are also more enriched than the modules produced by already benchmark frameworks. 

%Notice that if two group of genes $A$ and $B$ are both associated with a GO term and are put in different modules, then the corresponding GO term will be reported for both modules, but their odds ratios will be modest; possibly higher than 1, but not much. On the other hand, if $A$ and $B$ are placed in the same module, then the GO term will be reported once, but with a much higher oddsRatio (and p-value). This overall suggests that pSGC ans SGC tend to put into the same module groups of genes that are reported to belong to different modules by the other three methods, due to the limitations of the clustering methods used in these frameworks. In that way, SGC appears to report less terms with an odds ratio higher than 1, but that is due to the fact that 

%The red stack shows the proportion of the GO terms that have {\em odds ratios} less or equal to \( 1 \). Almost in all the cases pSGC and SGC have higher percentage in this bin relative to the other three frameworks; as expected, this is reversed for genes with {\em odds ratio} is greater than \( 1 \). This overall shows that the other frameworks have a tendency to find clusters with a modestly good odds ratio, which however remains close to 1. 

%As it seen, other pipelines than pSGC, and SGC have larger portion for this bin. Finally, the orange stack indicate the positive infinity {\em odds ratios}. 

\begin{figure}[!htb]
\centering
\begin{subfigure}{.48\textwidth}
  \centering
  \includegraphics[width=0.95\linewidth]{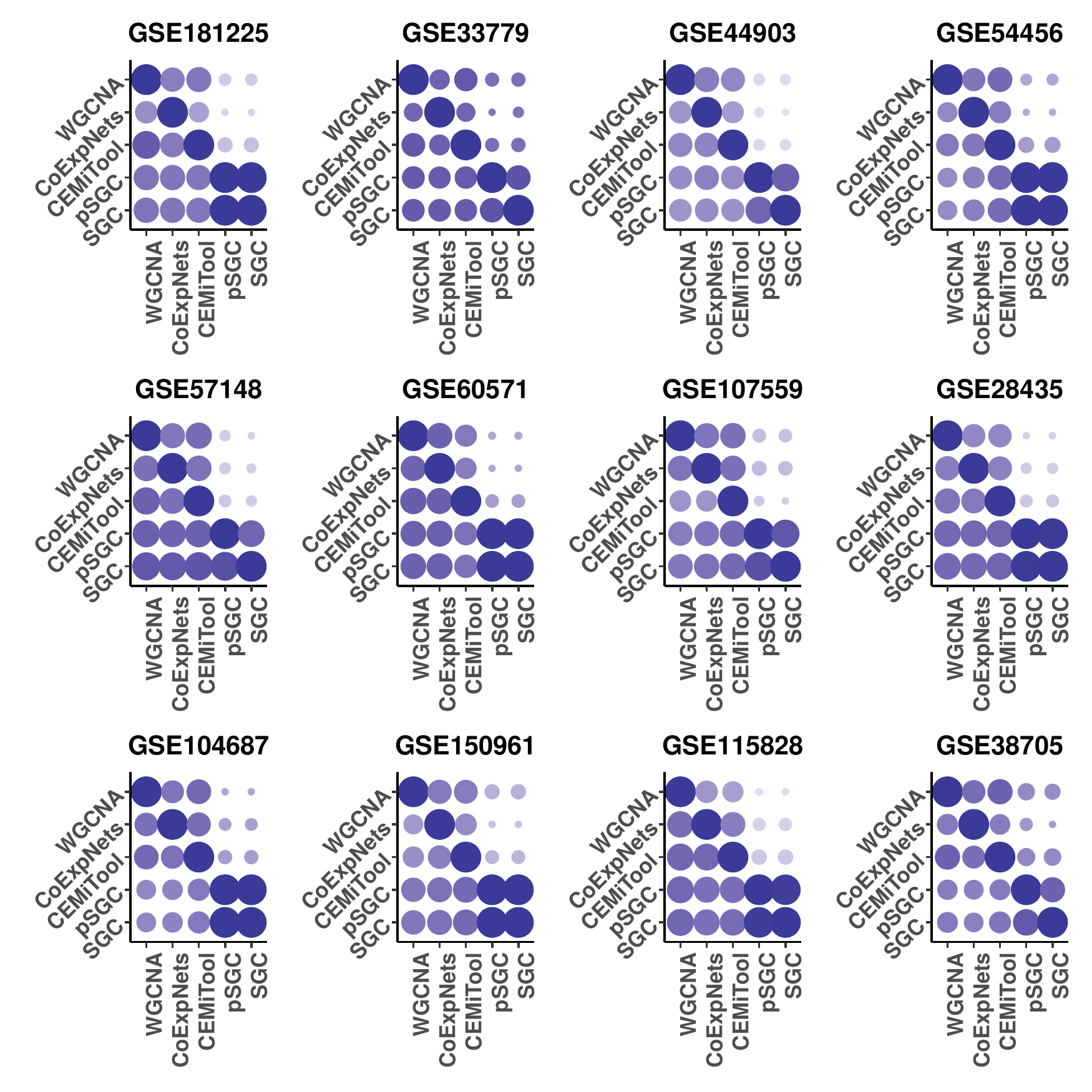}
  \caption{}
  \label{fig:strikeWhole}
\end{subfigure}
\begin{subfigure}{.48\textwidth}
  \centering
  \includegraphics[width=0.95\linewidth]{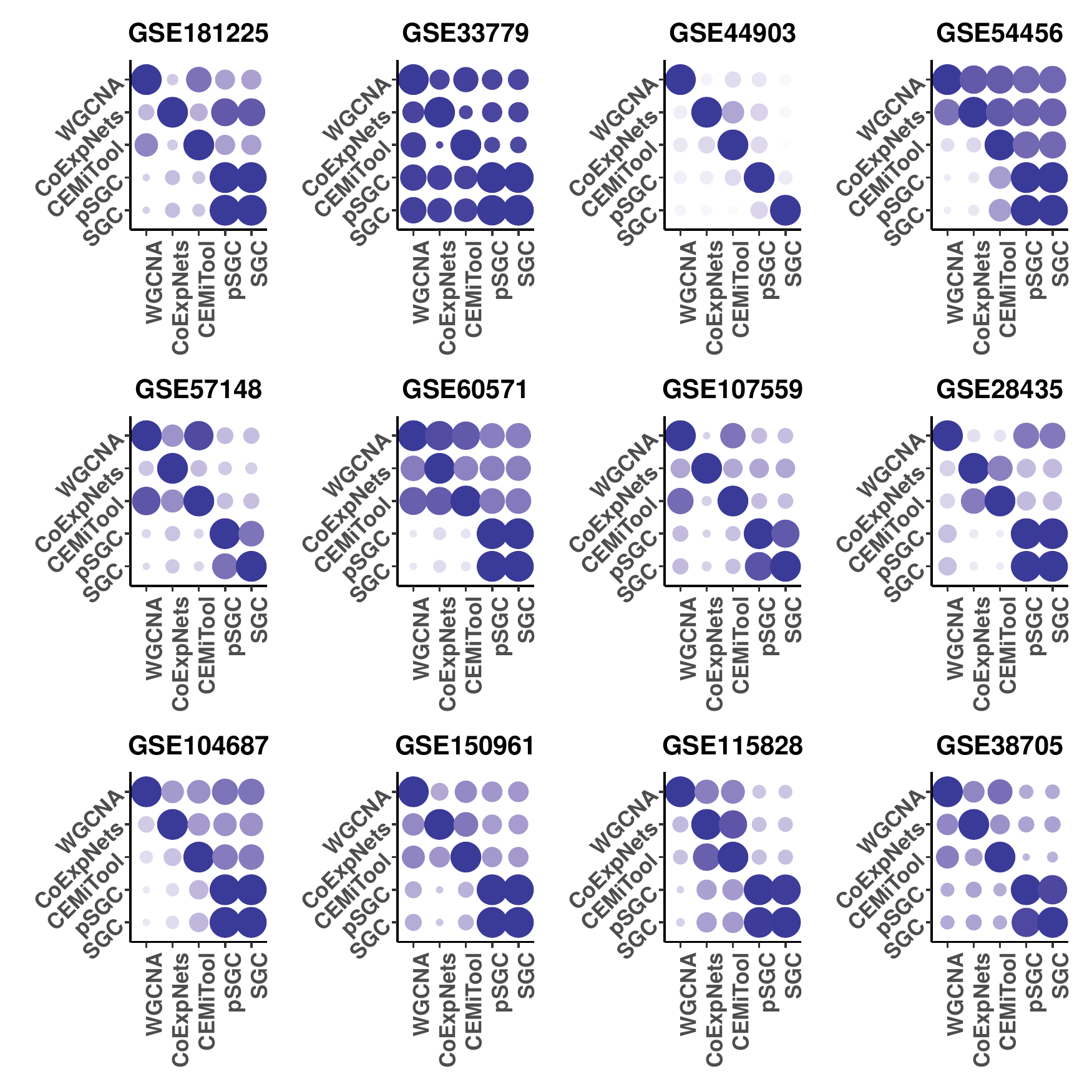}
  \caption{}
  \label{fig:strickProm}
\end{subfigure}
\caption{Overlaps GO terms reported by WGCNA, CoExpNets, CEMiTool, pSGC, and SGC in \( 12 \) real datasets. In both Figures (a) and (b), for pipeline $p$ in the $x$-axis and pipeline $q$ in the $y$-axis, position $(p,q)$ shows the number of unique GO terms reported by both $p$ and $q$, divided by the number of terms reported by $q$. Figure (a) accounts for GO terms reported in all modules, while in Figure (b) accounts for GO terms reported in the prominent (most significant) modules. The bigger and darker a circle the higher the percentage. } \label{fig:strike}
\end{figure}

\subsection{Observations} 

\par \textbf{Overlap in significant GO terms.}  It is interesting to investigate the overlap of GO terms reported by the different frameworks. To this end, we report two different measures for overlap, in~\autoref{fig:strike}. 
Not surprisingly, pSGC and SGC show significant overlaps with each other, as is the case with WGCNA, CEMiTool and CoExpNets, that also share algorithmic components. The overlaps between SGC and the other three frameworks are smaller, indicating that SGC reports GO terms that are not reported by the other frameworks. This is even more prominent when focusing in the most significant GO terms, reported in Figure~\ref{fig:OddsRatioPlot}.

\par It is also interesting to consider the information that is implied by the different sizes of the circles in symmetric positions $(p,q)$ and $(q,p)$. In Figure~\ref{fig:strike}(a), the circles in the lower-triangular part of the 12 arrays are larger than their symmetric counterparts. This shows that when considering all modules the number of GO terms reported by SGC is smaller. On the other hand, when considering the prominent module (defined as the module containing the GO term with most significant p-value), Figure~\ref{fig:strike}(b) shows that the prominent module for SGC contains a higher number of statistically significant GO terms, relative to the prominent module of other methods. Overall, these two observations show that SGC produces a prominent module with a higher density of statistically significant GO terms.

\begin{figure}[!htb]
\centering
\begin{subfigure}{.49\textwidth}
  \centering
  \includegraphics[width=.90\linewidth]{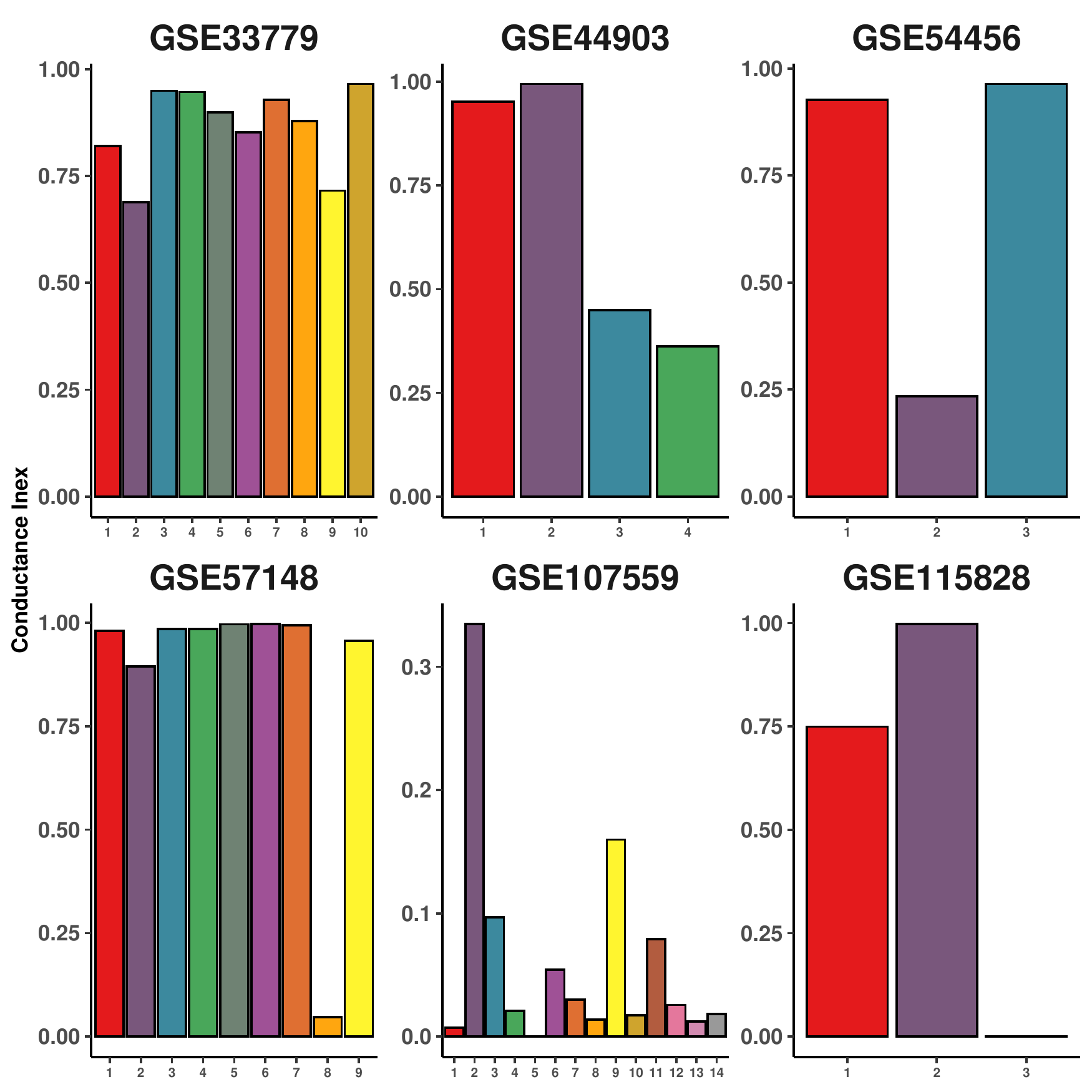}
  \caption{}
  \label{fig:conductance}
\end{subfigure}
\begin{subfigure}{.49\textwidth}
  \centering
  \includegraphics[width=.90\linewidth]{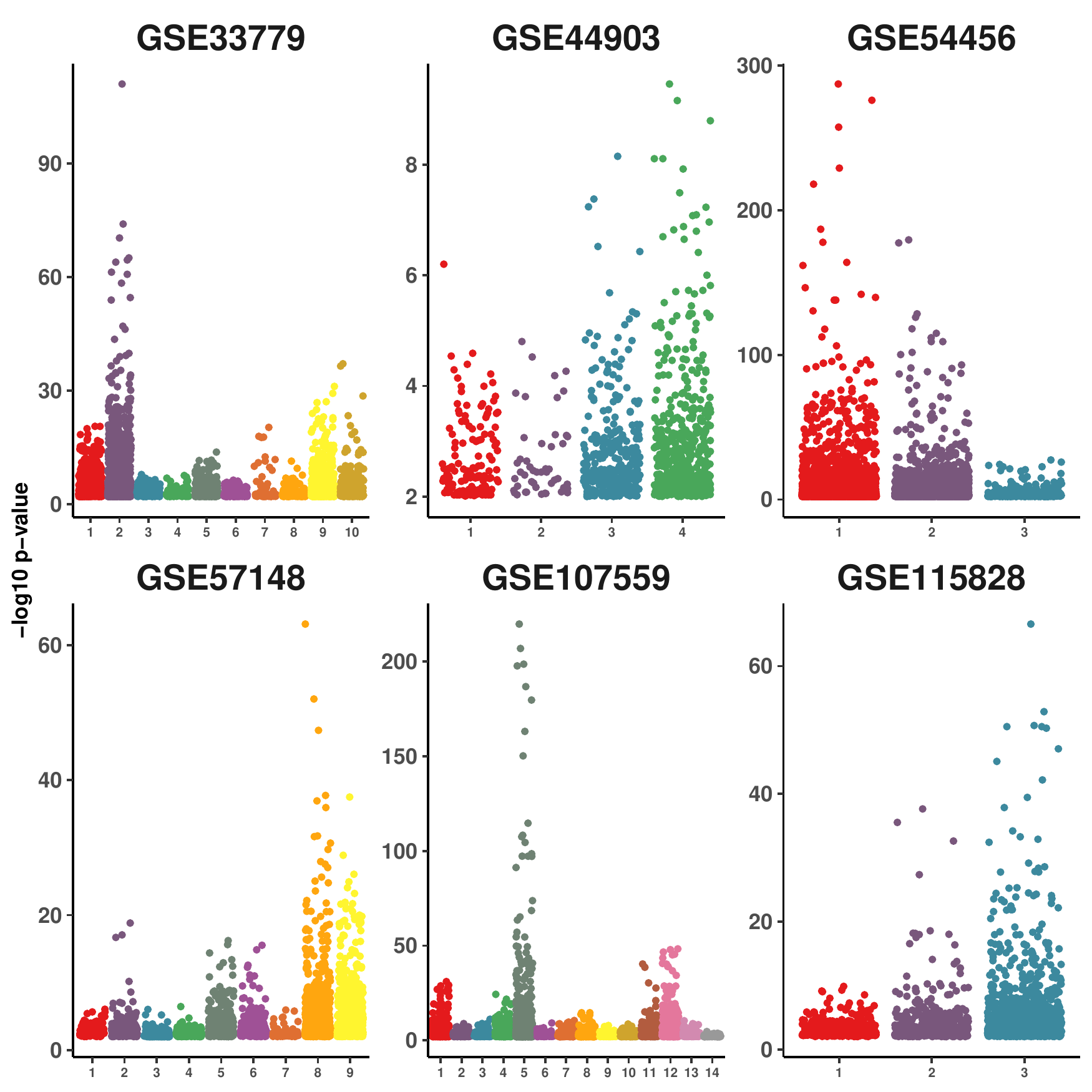}
  \caption{}
  \label{fig:alljitter}
\end{subfigure}
\caption{Conductance index and log-transformed {\em p-values} analysis in \( 6 \) real dataset. For each data, the conductance index for the clusters (on the left) along with its corresponding log-transformed {\em p-values} distribution (on the right) is depicted. (a) Conductance index for each module per data. The smaller the bar, the better the cluster. (b) log-transformed {\em p-values} for each module per data. The higher the point, the more enriched the GO term.}
\label{fig:condcutJitter}
\end{figure}
\begin{figure}[!htb]
\centering
\begin{subfigure}{.48\textwidth}
  \centering
  \includegraphics[width=.9\linewidth]{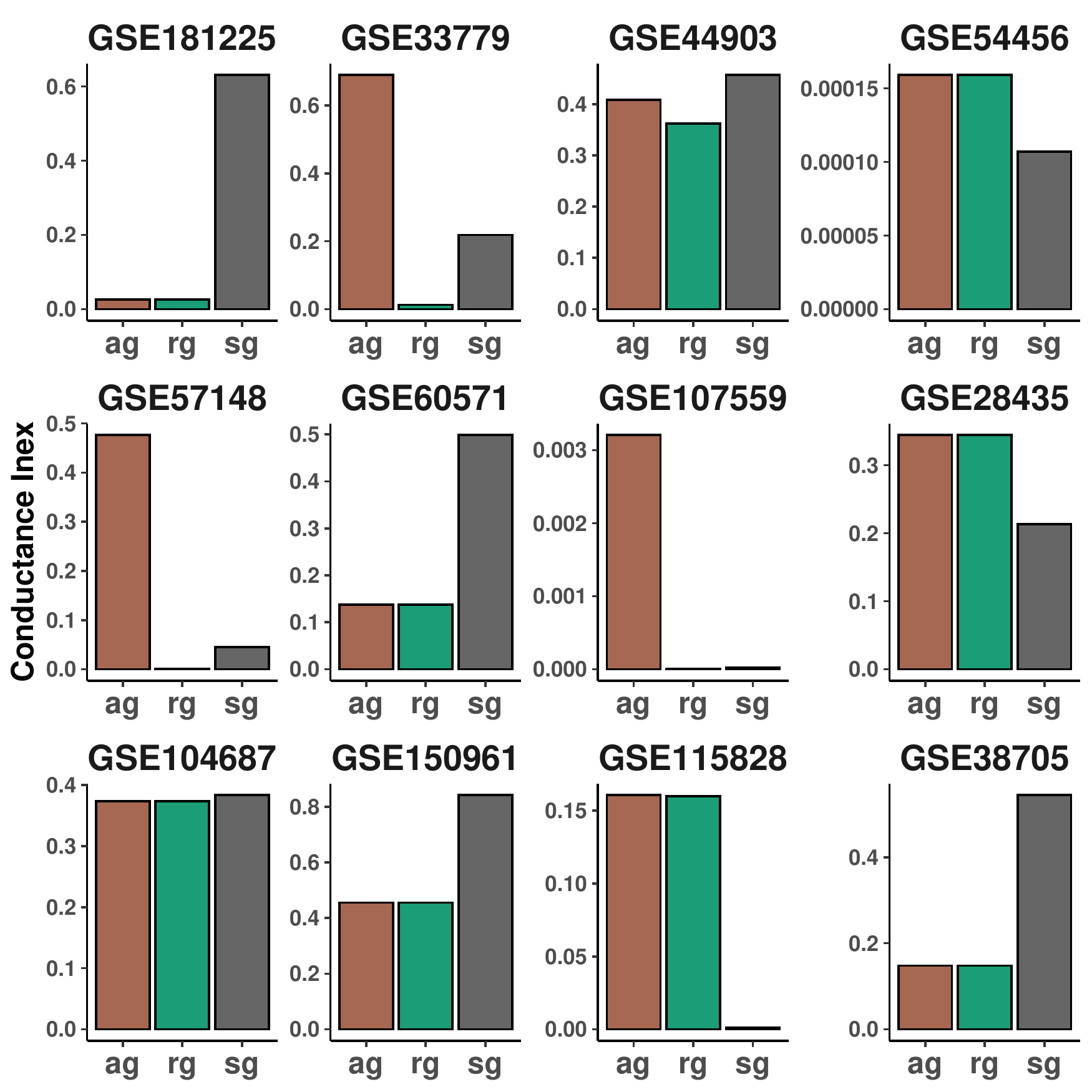}
  \caption{}
  \label{fig:fsgConduct}
\end{subfigure}
\begin{subfigure}{.48\textwidth}
  \centering
  \includegraphics[width=.9\linewidth]{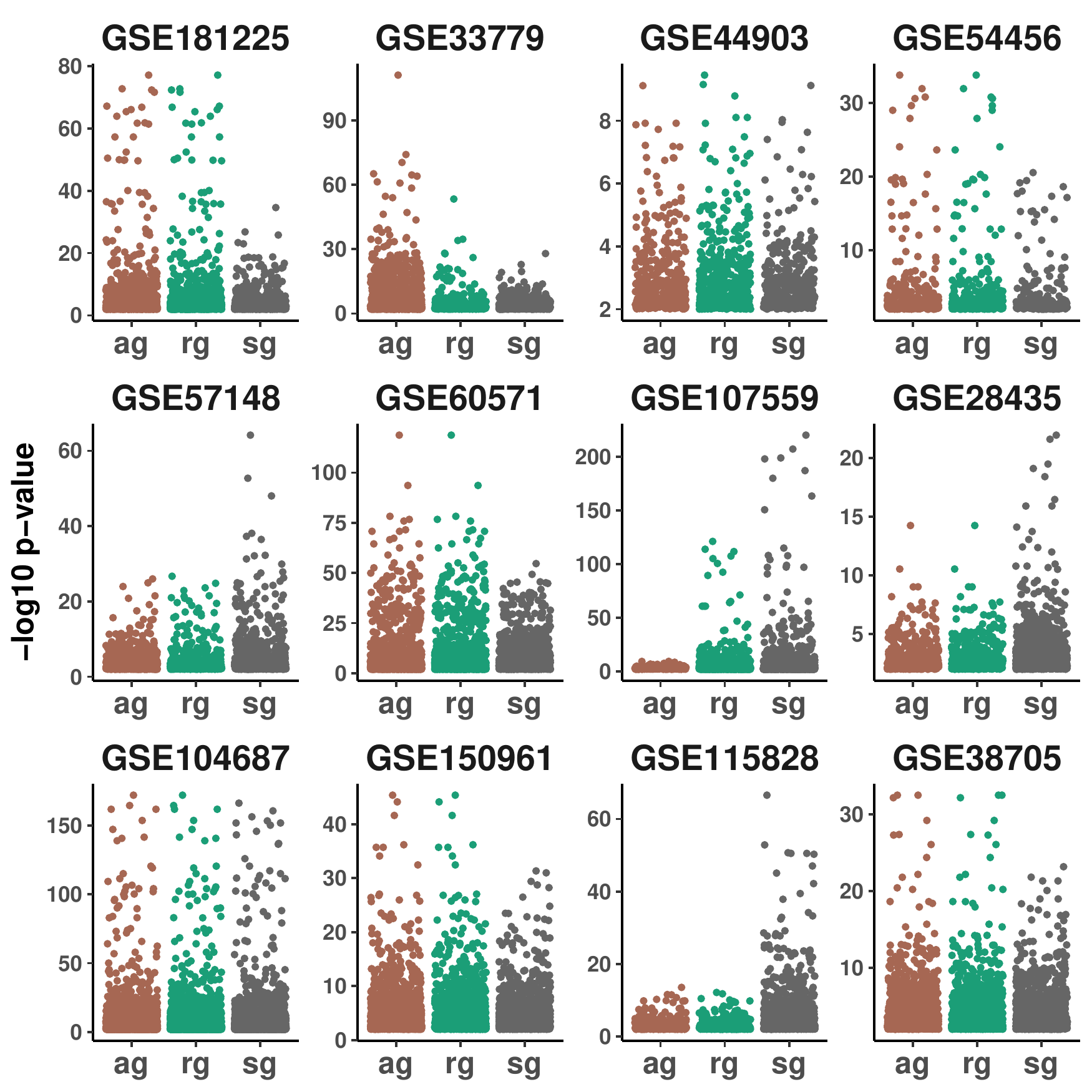}
  \caption{}
  \label{fig:fsgJitter}
\end{subfigure}
\caption{Conductance index and log-transformed {\em p-values} analysis for additive gap (``ag''), relative gap (``sg''), and second-order gap (``sg'') clusters in \( 12 \) real datasets. 
(b) log-transformed {\em p-values} of the selected clusters for ``ag", ``rg", and ``sg" are shown. The higher the point, the more significant the GO term.\label{fig:fsg}}
\end{figure}

\par {\bf The conductance measure.}
Spectral clustering targets the computation of clusters with a small conductance index as defined in \autoref{sec:conductance}~\cite{Cheeger}. Thus, when optimizing for conductance, we implicitly hypothesize that smaller conductance should correspond to higher module enrichment. 

We indeed have observed that there is correspondence between the cluster conductance index and cluster enrichment. \autoref{fig:condcutJitter} shows the conductance index of the modules computed by SGC, along with their their corresponding enrichment; here we focus in the cases when $k>2$. It can be seen that in all \( 6 \) data except GSE54456, clusters with smaller conductance indices have higher enrichment. In particular, in GSE107559, the modules with smaller conductance index were in order the clusters with label \( 5, 1, 13, 8, 10, 14, 4, 13\)~(see~\autoref{fig:conductance}). Interestingly, from~\autoref{fig:alljitter}, it can be seen that these clusters have higher enrichment.

As discussed in Section~\ref{sec:ourFramework} our framework relies on this connection of cluster conductance with enrichment to automatically compute a value of $k$ before computing the final clustering and the GO enrichment for the modules. In particular the method computes the enrichment of three test clusters, that were picked based on their conductance. These clusters' conductance and enrichment are reported in Figure~\ref{fig:fsg}, where the general correlation between conductance and enrichment is evident.

\begin{figure}[!htb]
    \centering
    \includegraphics[width=.75\linewidth]{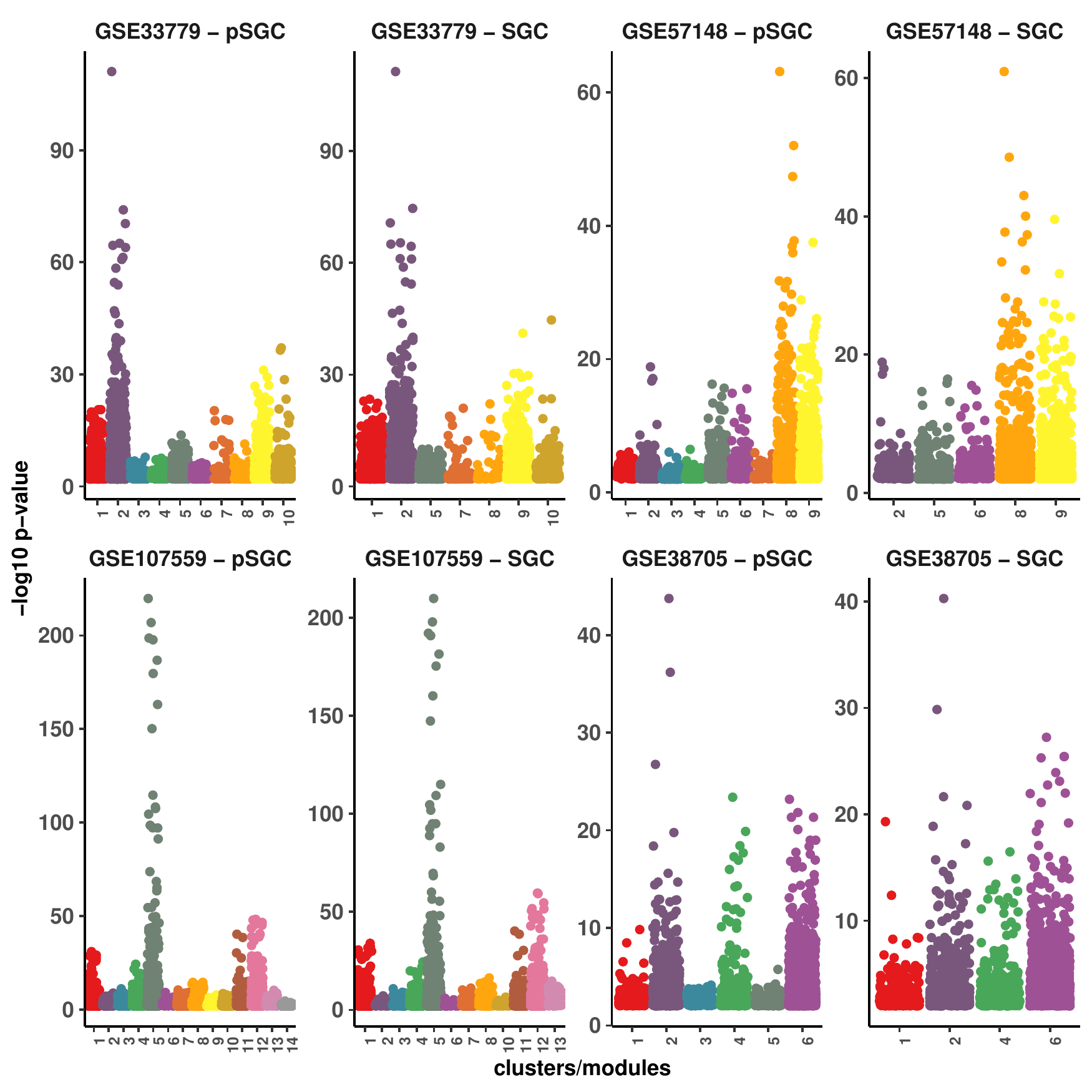}
    \caption{Comparing pSGC clusters and SGC modules in \( 4 \) real datasets. In all cases re-classification has resulted in a smaller number of modules relative to clusters. The labels of the eliminated clusters are \( 3, 4, 6 \), in GSE33779, \( 1, 3, 4, 10 \) in GSE57148, \( 9, 14 \) in GSE107559, and \( 3, 5 \) in GSE38705.}
    \label{fig:seljitter}
\end{figure}

\par {\bf Semi-supervised re-classification.} Once initial clusters by kmeans are produced, SGC carries out an additional semi-supervised re-classification of genes to return final modules, as described in~\autoref{sec:ourFramework}. A summary of the impact of this final step is given in~\autoref{tab:GSESummary} in the SGC column. ``\%UNR Genes" indicates the percentage of the total genes that are {\em unremarkable}, and ``\% CH Label" specifies the percentage of unremarkable genes whose label changed after the re-classification. Generally, when the percentage of unremarkable genes is small, the final modules agree with pSGC clusters; this happens in GSE104687, GSE181225, GSE54456, GSE107559, and GSE150961.  In contrast, for a higher percentage unremarkable genes, SGC assigns new labels to unremarkable genes and changes significantly the clusters' shape and size. The highest unremarkable gene percentages occurred in GSE33779, GSE57148, and GSE38705. The difference in enrichment between the clusters (pSGC) and modules (SGC) for these data is shown  in~\autoref{fig:seljitter}. It can be seen that, in all cases, the number of clusters gets reduced and the overall enrichment of the modules increases. In GSE107559 the percentage of unremarkable genes is relatively low, but re-classification has wiped out \( 2 \) clusters. In general, if there are clusters that are not enriched the re-classification step eliminates these clusters.

\section{Conclusion} \label{sec:conclusion}
\par We have proposed a novel framework for gene co-expression network analysis. Our pipeline  integrates multiple novel elements that deviate from existing frameworks in various ways. The GCN construction relies on a similarity measure that unlike previous works does not take into account network scale-freeness. The clustering algorithm departs from the currently used paradigm of Hierachical Clustering, and makes use of recent  progress in spectral clustering. The framework also includes a novel semi-supervised re-clustering step that takes into account gene ontology information unlike previous frameworks that include it only as an end result for module evaluation. We also make the observation that the conductance index, i.e. the optimization objective of spectral clustering is empirically correlated with high module enrichment, providing experimental support to our choice of spectral clustering. Our framework produces modules that have been found to be of superior quality on a comprehensive set of experiments with real datasets.

\newpage

%%%%%%%%%%%%%%%%%%%%%%%%%%%%%%%%%%%%%%%%%%%%%%%%%%%%%%%%%%%%%%%%%%%%%%%%%%%%%%%%%%%%%%%%%%%%%%%%%%%%%%%%%%%%%%%%%%
\subsection{Online Methods}
\label{subsec:onlineMethods}

\par \textbf{SGC framework.} SGC is a framework for the construction and analysis of Gene Co-expression Networks.

\smallskip
\noindent The input is a matrix \( GE_{m \times n} \) containing the gene expressions. In GE, rows and columns correspond to genes and samples respectively. Each entry \( ge_{i, j} \) is an expression value for gene \( i \) in sample \( j \). SGC does not perform any normalization or correction for batch effects and it is assumed that these preprocessing steps have been already performed. SGC is based on \( 5 \) main steps. Each step offers parameters that can be adjusted by the user.

\subsubsection{Step I: Network Construction}
\label{subsubsec:NetworkConstruction}

\paragraph{Gene-level Normalization.} In this step, each gene expression vector, i.e.~each row of the matrix \( GE_{m \times n} \) is divided by its Euclidean norm which is calculated as
\begin{equation}
\label{eq:EculideanNorm}
\lVert GE_{i,.} \rVert_{2} =  \sqrt{ge^2_{i,1}, \dots, ge^2_{i,n}},
\end{equation}

\noindent  where \( GE_{i,.} = <ge_{i,1}, \dots, ge_{i,n}> \) is the expression vector of gene $ i $. The result of this step is matrix \( G_{m \times n} \).

\paragraph{Similarity Calculation.} We calculate the variance $\gamma^2$ over all $ m^2/2 $ pairwise Euclidean distances $ \lVert g_{i}-g_{j} \rVert_{2}^2 $. We then calculate the following type of Gaussian kernel for each pair of genes. 

\begin{equation}
    \label{eq:GaussianKernel}
     s_{i,j} = k(g_{i},g_{j}) = \exp(\frac{-\lVert g_{i}-g_{j}\rVert_{2}^2}{2 \gamma^2}).
\end{equation}
The result is a similarity matrix \( S_{m \times m} \) where \( m \) is the number of the genes. Note that \( S \) is a symmetric square matrix which ranges from \( 0 \) for the most dissimilar to \( 1 \) for the most similar genes.

\paragraph{Topological Overlap Enhancement.} The adjacency of the network is derived by adding second-order neighborhood information to \( S_{m \times m} \) in the form of the topological overlap measure (TOM)~\cite{Zhang2005general, Lan2008}. The adjacency strength between gene $i$ and $j$ is calculated by the following formula:

\begin{equation}
\label{eq:TOM}
a_{i,j} = \frac{l_{i,j} + s_{i,j}}{\min{(k_i,k_j)} + 1 - s_{i,j}},
\end{equation}

\noindent  where $ l_{i,j}  = \sum_{u}s_{i,u}s_{u,j} $, and $ s_{i,j} $ is the similarity coefficient between gene \( i \) and \( j \) from matrix \( S \) of the previous step, and $k_i = \sum_{j}s_{ij} $ is the degree of node $ i $. The output is a symmetric adjacency matrix \( A_{m \times m} \) with values in \( [0,1] \) where \( m \) is the number of genes. Note that the diagonal elements of \( A \) is zero.

%\par We also propose a novel measure call \textbf{\textit{Deep Overlap Measure}} (DOM). Similar to TOM, DOM captures and adds the information of third order neighborhoods of the genes to the network in form of~\autoref{eq:DOM}
%\begin{equation}
%\label{eq:DOM}
%\phi_{i,j} = \frac{s_{i,j} + l_{i,j} + t_{i,j}}{\min{(k_i,k_j)} + \min{(f_i, f_j)} + 1 - %s_{i,j}}
%\end{equation}

%\par where \( t_{i,j} = \sum_{u,v}s_{i,u}s_{u,v}s_{v,j} \), and \( f_{i}  = \sum_{u}s^{2}_{iu}\). There rest of parameters are the same as TOM~(i.e. ~\autoref{eq:TOM}). By default, SGC calculates TOM, however, users can specify DOM by setting \textit{flagDOM} = TRUE. 

%\par The result of this step, is a squared symmetric adjacency matrix \( A_{m \times m} \) with values in \( [0,1] \) where \( m \) is the number of genes. Note that diagonal elements of \( A \) is zero.

\subsubsection{Step II: Network Clustering}
\label{subsubsec:networkClustering}
\paragraph{Eigenvalues and Eigenvectors.} Let \( A \) be the adjacency matrix from the previous step. Let $D$ be the diagonal matrix containing the degrees of the nodes in the similarity matrix, i.e. $d_{ii} = \sum_{j} a_{ij}$. We perform the following steps: 

\begin{itemize}
     \item Compute the eigenvalues and the corresponding eigenvectors of $D^{-1}A$.  Let $\lambda_1,\ldots,\lambda_m$ be the eigenvalues, and $Y_1,\ldots,Y_m$ be the corresponding eigenvectors. 
     \footnote{The number of eigenvectors that is practically needed for the rest of the pipeline never exceeds $m'=50$. With an appropriate method one can calculate at most $m'$ eigenvectors, resulting in a faster method. } 
     \item For eigenvector $Y_i$ define the scalar $a_i = {\bf 1}^T D Y_i/m$, where $\bf 1$ is the all-ones vector. Then subtract $t_i$ from each entry of $Y_i.$
     \item Let $Y_i:= Y_i/(Y_i^T D Y_i)$.
     \item Drop the first column of $ Y $.
\end{itemize}

\par The output of this step consists of the eigenvalues $\lambda_1,\ldots,\lambda_m$, and of the matrix of eigenvectors $Y_{m - 1 \times m}$, where eigenvector $Y_i$ is the $i^{th}$ column of $Y$.

%\begin{enumerate}[label=\Roman*]
%    \item \( Y = D^{-1/2} \cdot Y \).
%    \item \( u = (v^t \cdot Y) \) where \( v^t \) is the transpose of vector \( v \).
%    \item \( u = u \div s \)  where \( s = \sum_{i=1}^{m} v_{i} \) and \( \div \) is the division of every element in \( u \) by \( s \).
%    \item \( y = y/c \) where \( c = y^t \cdot D^{-1/2} \cdot y \), for every column \( y \) in \( Y \).
%\end{enumerate}

 %\par The result is a squared matrix \( Y_{m \times m} \) where \( m \) is the number of the genes. Since the first eigenvalue of a symmetric squared matrix \( A \) is always \( 1 \), the corresponding column in \( Y \) is dropped resulting in \( Y_{m \times m-1} \) matrix .

\paragraph{Determining the Number of Clusters.} Three potentially different values for the number of clusters are calculated, $kag,krg,ksg$ using respectively what we call the \textit{additive gap}, \textit{relative gap} and the \textit{second-order gap} methods. These are calculated as follows:

\begin{equation}
\label{eq:kgap}
kag = \argmax_i \left(\lambda_{i + 1} - \lambda_{i} \right) \qquad \textrm{for} \quad i = 2, \dots, m-1
\end{equation}

\begin{equation}
\label{eq:kfirst}
    krg = \argmax_i \left( \frac{1- \lambda_{i+1}}{1- \lambda_{i}} \right) \qquad \textrm{for} \quad i = 2, \dots, m-2
\end{equation}

%\begin{equation}
%    \label{eq:ksecond}
%    ksg = \argmax_i \left(f_{i} - f_{i+1}\right) \qquad  \textrm{for} \quad i = 2, \dots, m
%\end{equation}

\begin{equation}
    \label{eq:ksecond}
    ksg = \argmax_i \left(\frac{1- \lambda_{i+1}}{1- \lambda_{i}} - \frac{1- \lambda_{i+2}}{1- \lambda_{i+1}}\right) \qquad  \textrm{for} \quad i = 2, \dots, m
\end{equation}

%\par Then, the number of clusters in additive gap, relative gap and second-order gap are as follows.

%\begin{equation}
%\label{eq:kgap}
%kag = \arg \max_{i} ag_i
%\end{equation}

%\begin{equation}
%\label{eq:kfirt}
%krg = \arg \max_{i} rg_{i}
%\end{equation}

%\begin{equation}
%\label{eq:ksecond}
%ksg = \arg \max_{i} sg_{i}
%\end{equation}

%\par Here $\lambda_1,\ldots,\lambda_m$ are the eigenvalues. %Note that \( \lambda_1\) is dropped since the first eigenvalue in \( X \) is always \( 1 \).

%\par And \( kag \), \( krg \), and \( ksg \) denote the number of clusters for methods additive gap, relative gap, and second-order gap respectively.  

%\Yiannis{Below: We should let $Y'$ be the first $2k+1$ columns, unless the first column has been dropped from $Y$}

\paragraph{Calculation of Conductance Index.}\label{sec:conductance} For each of the three possible values of \( k \) (i.e. $kag,krg,ksg)$, We set $Y'$ to consist of the \( 2k \) columns (i.e.~eigenvectors) of \( Y \). Each row in \( Y^{\prime} \) is then divided by its Euclidean norm %~\autoref{eq:EculideanNorm} 
so that length of each row becomes \( 1 \). Next, the kmeans clustering algorithm~\cite{KMeans} is applied on \( Y^{\prime} \) to find \( k \) clusters using the default kmeans() R function. By default, the maximum number of iterations is set to \( 10^{8} \) and the number of starts is set to \( 1000 \). Then, for each cluster the conductance index is computed. Let \( C_i \) be one of the clusters. The conductance index for cluster \( C_i \) is defined in~\autoref{eq:conductance}. 

%\Yiannis{In the results/discussion part, you refer to conductance. People will not know it, so you have to put a reference to this equation}

\begin{equation}
\label{eq:conductance}
conduct{(C_i)} = \frac{ \sum_{u \in C_i,v \not \in C_i} a_{u,v}}{\sum_{u \in C_i}deg(u)}
\end{equation}
where \( deg(u) = \sum_{j} A_{u, j}\) which indicates the degree node \( u \) (sum of all the weights associated to node \( u \)), and \( a_{u,v} \) is the pairwise association between node \( u \) and \( v \) in adjacency matrix \( A \). For each method, the cluster that has the minimum conductance index is chosen and passed to the next level. Let \( c_{ag} \), \( c_{rg} \), and \( c_{sg} \) denote the clusters with minimum conductance index for the three aforementioned methods respectively.

\paragraph{Gene Ontology Validation.}

\par In this step, the enrichment of clusters \( c_{ag} \), \( c_{rg} \), and \( c_{sg} \) are calculated using the GOstats~\cite{GOstats} R package individually for all six possible queries (``underBP",~``overBP",~``underCC",~``overCC",~``underMF",~``overMF`") combined. To this end, a conditional ``hyperGTest" test is performed and the entire set of genes in the data is considered for the ``universeGeneIds". For each cluster \( c \in \{c_{ag}, c_{rg}, c_{sg}\} \), GOstats returns the GO terms found in \( c \) along with a {\em p-value} for each term. Let \( P_{i} \) denote the {\em p-value} associated with a GO term $i$ found in \( c \).  Then the quality
of a cluster $c$ is determined by:
\begin{equation}
    \sum_{j \in c} -\log_{10}(P_j).
\end{equation}
This measure is then used to pick the cluster of best quality among $\{c_{ag}, c_{rg}, c_{sg}\}$. Each of these three clusters was produced by kmeans with a specific choice of $k$: $kag,krg,ksg$ respectively. Then the cluster of best quality directly determines what value of $k$ will be used. For example if $c_{ag}$ is the best cluster, then $k=k{ag}$. After determining $k$, the clusters computed earlier by kmeans for that value of $k$ are returned as output, along with embedding matrix \( Y^{\prime}_{m \times 2k} \).

%Then the best among the three methods is is determined as follows:
%\begin{equation}
%\label{eq:finalk}
% \arg \max_{a, r, g} \{ \sum_{j \in g} -\log_{10}(P_j), \sum_{j \in f} -\log_{10}(P_j), \sum_{j \in s} -\log_{10}(P_j) \}
%\end{equation}

%\par Let \( k \) is the cluster number correspond to winner method \( i \). At the end of this step, the clusters produced by method \( i \) along with embedded matrix \( Y^{\prime}_{m \times 2*k} \) (the first \( 2*k \) columns of \( Y \)) is passed to the next step.

\subsubsection{Gene Ontology}
\label{sec:geneOntology}

\par The GOstats R package~\cite{GOstats} is applied to each cluster returned in the GO Validation step. The settings of GOstats are the same as in the GO validation step.  GOstats reports answers on user-specified queries including ``id", ``term", ``p-value", ``odds" ``ratio", ``expected count", ``count", ``size". SGC reports this information for each cluster separately. Additionally, for each cluster SGC reports the GO terms that have been found in the cluster.

\subsubsection{Gene Semi-Labeling}
\label{sec:SemiLabeling}

\par In the default setting, SGC picks the top \( 10 \% \) GO terms according to their associated {\em p-values}, and consider their corresponding genes as {\em remarkable}. All other genes are considered {\em unremarkable}. That percentage is user-adjustable.

\par  With this definition, some clusters may not contain any remarkable genes. Then, each remarkable gene inherits the label of its parent cluster. The unremarkable genes remain unlabeled.

\subsubsection{Supervised Classification}
\label{sec:SemiSupervisedClassification}

\par Labeled and unlabeled gene sets along with their corresponding $2k$-dimensional points given by the rows of \( Y^{\prime} \) (obtained in the Network clustering step) define a semi-supervised classification problem. We adopt a simple solution that uses the embeddings of the labeled genes as training points, and we train a simple classifier such as \textit{k-nearest neighbors} (kNN)~\cite{knn1,knn2} or logistic regression~\cite{lr}. Then the trained classifier is used to classify the unlabeled points, and their corresponding genes. Note that $k$ is the number of clusters determined in the Network Clustering step, but the actual number of clusters returned in this step is equal to the number of clusters found to contain remarkable genes in the {\em Gene Semi-labeling} step. The default model is kNN and the number of neighbors is ranging from \( 20:(20 + 2*k) \) if \( 2*k \leq 30\) otherwise \( 20:30 \) depending on accuracy metric using~\cite{caret} R-package.

%are used to obtain labels for the unlab. The model is called \textit{semi-supervised} since all the datapoints are not labeled, and it makes prediction of unlabeled genes on the basis of the labeled one. Note that for each gene, the length of its corresponding vector is \( 2*k \) where \( k \) is the optimal number of cluster found in step Network Clustering. After this step, genes that belong to the same class form a module and number of modules are equal to number of clusters that remarkable at least one GO term has been found in it. 

\subsection{Final remark.} The proposed pipeline is flexible. In particular, SGC enables the user to define their preferred number f clusters $k$. For example, we have found that in the case of dataset GSE44903, SGC outperforms the baselines significantly with a different value of $k$ that is not automatically produced by our pipeline. SGC also includes a user-defined threshold about the percentage of GO terms used for finding remarkable genes and clusters.

\subsection{Settings in baseline pipelines}
\label{sec:pipelines}

%\par As discussed earlier, all the pipelines in the benchmark use soft-power (sft) to make the GCNs scale-free. For each data, same soft-power is used for scale-free constraint and is written in~\autoref{tab:GSESummary}. Function ``blockwiseModules" is used for ``signed" GCNs construction and produce modules for WGCNA. Function ``getDownstreamNetwork" is used for ``signed" GCNs construction and analysis in CoExpNets. Finally, ``cemitool" function is used to build ``signed" GCNs construction and produce modules for CEMiTool framework.

\par As discussed earlier, all the baseline pipelines use soft-powering (sft) to make the GCNs scale-free. We use the same soft-power methods across all pipelines and the specifics powers used for each dataset are reported in~\autoref{tab:GSESummary}. The functions that are used for GCN construction and analysis in WGCNA, CoExpNets and CEMiToo, are ``blockwiseModules'', ``getDownstreamNetwork" and ``cemitool'' respectively. 

%We use the same soft-powering method to determine the powers, and the specifics powers used for each dataset are reported in~\autoref{tab:GSESummary}. The functions that are used for GCN construction and analysis in WGCNA, CoExpNets and CEMiToo, are ``blockwiseModules'', ``getDownstreamNetwork" and ``cemitool'' respectively

\section{Supplementary Files} 
\label{sec:supp}

%\Yiannis{Please link the supplementary files here}

\par \href{run:./SupplementaryFile1_PCA.xlsx}{SupplementaryFile1\_PCA}

\par \href{run:./SupplementaryFile1_HeatMaps.xlsx}{SupplementaryFile2\_HeatMaps}

\par \href{run:./SupplementaryFile1_PipelineModuleDetail.xlsx}{SupplementaryFile2\_PipelineModuleDetail}

\par \href{run:./SupplementaryFile1_GOtermDistributions.xlsx}{SupplementaryFile2\_GOtermDistributions}

\bibliographystyle{plain}
\bibliography{reference.bib}

\begin{thebibliography}{10}

\bibitem{Abbas-Aghababazadeh2018}
Farnoosh Abbas-Aghababazadeh, Qian Li, and Brooke~L. Fridley.
\newblock Comparison of normalization approaches for gene expression studies
  completed with high-throughput sequencing.
\newblock {\em PloS one}, 13(10):e0206312--e0206312, Oct 2018.

\bibitem{niloo2021}
Niloofar Aghaieabiane and Ioannis Koutis.
\newblock A {N}ovel {C}alibration {S}tep in {G}ene {C}o-{E}xpression {N}etwork
  {C}onstruction.
\newblock {\em Frontiers in Bioinformatics}, 1, 2021.

\bibitem{knn2}
N.~S. Altman.
\newblock An introduction to {k}ernel and {N}earest-{N}eighbor {N}onparametric
  regression.
\newblock {\em The American Statistician}, 46(3):175--185, 1992.

\bibitem{GEO}
Tanya Barrett, Stephen~E. Wilhite, Pierre Ledoux, Carlos Evangelista, Irene~F.
  Kim, Maxim Tomashevsky, Kimberly~A. Marshall, Katherine~H. Phillippy,
  Patti~M. Sherman, Michelle Holko, Andrey Yefanov, Hyeseung Lee, Naigong
  Zhang, Cynthia~L. Robertson, Nadezhda Serova, Sean Davis, and Alexandra
  Soboleva.
\newblock {NCBI} {GEO}: archive for functional genomics data sets--update.
\newblock {\em Nucleic acids research}, 41(Database issue):D991--D995, Jan
  2013.

\bibitem{GSE38705}
Brian~J Bennett, Charles~R Farber, Anatole Ghazalpour, Calvin Pan, Nam Che,
  Pingzi Wen, Hong~Xiu Qi, Adonisa Mutukulu, Nathan Siemers, Isaac Neuhaus,
  Roumyana Yordanova, Peter Gargalovic, Matteo Pellegrini, Todd Kirchgessner,
  and Aldons~J Lusis.
\newblock Unraveling inflammatory responses using systems genetics and
  gene-environment interactions in macrophages.
\newblock {\em Cell}, 151(3):658--670, 2012.

\bibitem{bishop}
Christopher~M. Bishop.
\newblock {\em Pattern Recognition and Machine Learning (Information Science
  and Statistics)}.
\newblock Springer-Verlag, Berlin, Heidelberg, 2006.

\bibitem{Botia2017}
Juan~A. Botía, Jana Vandrovcova, Paola Forabosco, Sebastian Guelfi, Karishma
  D’Sa, The United Kingdom Brain~Expression Consortium, John Hardy,
  Cathryn~M. Lewis, Mina Ryten, and Michael~E. Weale.
\newblock An additional k-means clustering step improves the biological
  features of {WGCNA} gene co-expression networks.
\newblock {\em BMC Systems Biology}, 11(1):47, 2017.

\bibitem{Broido2019}
Anna~D. Broido and Aaron Clauset.
\newblock Scale-free networks are rare.
\newblock {\em Nature communications}, 10(1):1017--1017, Mar 2019.

\bibitem{GSE60571}
Zhen-Xia Chen and Brian Oliver.
\newblock X {C}hromosome and {A}utosome {D}osage {R}esponses in {D}rosophila
  melanogaster heads.
\newblock {\em G3 (Bethesda, Md.)}, 5(6):1057--1063, Apr 2015.

\bibitem{Cheng2020}
Chew~Weng Cheng, David~J. Beech, and Stephen~B. Wheatcroft.
\newblock Advantages of {CEM}i{T}ool for gene co-expression analysis of
  {RNA}-seq data.
\newblock {\em Computers in Biology and Medicine}, 125:103975, 2020.

\bibitem{Clote2020}
P.~Clote.
\newblock Are {RNA} networks scale-free?
\newblock {\em Journal of mathematical biology}, 80(5):1291--1321, Apr 2020.

\bibitem{Emamjomeh2017}
Abbasali Emamjomeh, Elham Saboori~Robat, Javad Zahiri, Mahmood Solouki, and
  Pegah Khosravi.
\newblock Gene co-expression network reconstruction: a review on computational
  methods for inferring functional information from plant-based expression
  data.
\newblock {\em Plant Biotechnology Reports}, 11(2):71--86, 2017.

\bibitem{GOstats}
S.~Falcon and R.~Gentleman.
\newblock Using {GO}stats to test gene lists for go term association.
\newblock {\em Bioinformatics}, 23(2):257--258, 11 2006.

\bibitem{knn1}
Evelyn Fix and J.~L. Hodges.
\newblock Discriminatory {A}nalysis. {N}onparametric {D}iscrimination:
  {C}onsistency {P}roperties.
\newblock {\em International Statistical Review / Revue Internationale de
  Statistique}, 57(3):238--247, 1989.

\bibitem{Gat2003}
I~Gat-Viks, R~Sharan, and R~Shamir.
\newblock Scoring clustering solutions by their biological relevance.
\newblock {\em Bioinformatics}, 19(18):2381--9, 2003.

\bibitem{coseq}
Antoine Godichon-Baggioni, Cathy Maugis-Rabusseau, and Andrea Rau.
\newblock Clustering transformed compositional data using {K}-means, with
  applications in gene expression and bicycle sharing system data.
\newblock {\em Journal of Applied Statistics}, 46(1):47--65, 2019.

\bibitem{KMeans}
J.~A. Hartigan and M.~A. Wong.
\newblock Algorithm {AS} 136: A {K-Means} clustering algorithm.
\newblock {\em Applied Statistics}, 28(1):100--108, 1979.

\bibitem{GSE33779}
Raul Herranz, Oliver~J. Larkin, Richard~JA Hill, Irene Lopez-Vidriero, Jack~JWA
  van Loon, and F.~Javier Medina.
\newblock Suboptimal evolutionary novel environments promote singular altered
  gravity responses of transcriptome during {D}rosophilametamorphosis.
\newblock {\em BMC Evolutionary Biology}, 13(1):133, 2013.

\bibitem{Hou2021}
Jie Hou, Xiufen Ye, Chuanlong Li, and Yixing Wang.
\newblock K-{M}odule {A}lgorithm: {A}n {A}dditional {S}tep to {I}mprove the
  {C}lustering {R}esults of {WGCNA} {C}o-{E}xpression {N}etworks.
\newblock {\em Genes}, 12(1):87, 2021.

\bibitem{CCor2016}
Yiming Hu and Hongyu Zhao.
\newblock {CCor}: {A} whole genome network-based similarity measure between two
  genes.
\newblock {\em Biometrics}, 72(4):1216--1225, 2016.

\bibitem{RMA}
Rafael~A. Irizarry, Bridget Hobbs, Francois Collin, Yasmin~D. Beazer‐Barclay,
  Kristen~J. Antonellis, Uwe Scherf, and Terence~P. Speed.
\newblock Exploration, normalization, and summaries of high density
  oligonucleotide array probe level data.
\newblock {\em Biostatistics}, 4(2):249--264, 04 2003.

\bibitem{Raya2006}
Raya Khanin and Ernst Wit.
\newblock How {S}cale-{F}ree {A}re {B}iological {N}etworks.
\newblock {\em Journal of Computational Biology}, 13(3):810--818, 2006.

\bibitem{Khatri2005}
Purvesh Khatri and Sorin Drăghici.
\newblock Ontological analysis of gene expression data: current tools,
  limitations, and open problems.
\newblock {\em Bioinformatics}, 21(18):3587--3595, 2005.

\bibitem{GSE57148}
Woo~Jin Kim, Jae~Hyun Lim, Jae~Seung Lee, Sang-Do Lee, Ju~Han Kim, and Yeon-Mok
  Oh.
\newblock Comprehensive {A}nalysis of {T}ranscriptome {S}equencing {D}ata in
  the {L}ung {T}issues of {COPD} {S}ubjects.
\newblock {\em International Journal of Genomics}, 2015:206937, Mar 2015.

\bibitem{caret}
Max Kuhn.
\newblock {B}uilding {P}redictive {M}odels in {R} {U}sing the caret {P}ackage.
\newblock {\em Journal of Statistical Software}, 28(5):1–26, 2008.

\bibitem{Lan2008}
Peter Langfelder and Steve Horvath.
\newblock {WGCNA}: an {R} package for weighted correlation network analysis.
\newblock {\em BMC Bioinformatics}, 9(1):559, Dec 2008.

\bibitem{dynamicTreeCut}
Peter Langfelder, Bin Zhang, and Steve Horvath.
\newblock Defining clusters from a hierarchical cluster tree: the {D}ynamic
  {T}ree {C}ut package for {R}.
\newblock {\em Bioinformatics}, 24(5):719--720, 11 2007.

\bibitem{Cheeger}
James~R. Lee, Shayan~Oveis Gharan, and Luca Trevisan.
\newblock Multiway spectral partitioning and higher-order cheeger inequalities.
\newblock {\em J. ACM}, 61(6), dec 2014.

\bibitem{GSE54456}
Bingshan Li, Lam~C. Tsoi, William~R. Swindell, Johann~E. Gudjonsson, Trilokraj
  Tejasvi, Andrew Johnston, Jun Ding, Philip~E. Stuart, Xianying Xing, James~J.
  Kochkodan, John~J. Voorhees, Hyun~M. Kang, Rajan~P. Nair, Goncalo~R.
  Abecasis, and James~T. Elder.
\newblock Transcriptome analysis of psoriasis in a large case-control sample:
  {RNA}-seq provides insights into disease mechanisms.
\newblock {\em The Journal of investigative dermatology}, 134(7):1828--1838,
  Jul 2014.

\bibitem{Lima2009}
Gipsi Lima-Mendez and Jacques van Helden.
\newblock The powerful law of the power law and other myths in network biology.
\newblock {\em Mol. BioSyst.}, 5:1482--1493, 2009.

\bibitem{Liu2016}
Jing Liu, Ling Jing, and Xilin Tu.
\newblock Weighted gene co-expression network analysis identifies specific
  modules and hub genes related to coronary artery disease.
\newblock {\em BMC Cardiovascular Disorders}, 16(1):54, Mar 2016.

\bibitem{affymetrix}
D~J Lockhart, H~Dong, M~C Byrne, M~T Follettie, M~V Gallo, M~S Chee,
  M~Mittmann, C~Wang, M~Kobayashi, H~Horton, and E~L Brown.
\newblock Expression monitoring by hybridization to high-density
  oligonucleotide arrays.
\newblock {\em Nature biotechnology}, 14(13):1675--80, 1996.

\bibitem{Ma2018}
Xuelian Ma, Hansheng Zhao, Wenying Xu, Qi~You, Hengyu Yan, Zhimin Gao, and Zhen
  Su.
\newblock Co-expression {G}ene {N}etwork {A}nalysis and {F}unctional {M}odule
  {I}dentification in {B}amboo {G}rowth and {D}evelopment.
\newblock {\em Frontiers in Genetics}, 9:574, 2018.

\bibitem{McCall2010}
Matthew~N. McCall, Benjamin~M. Bolstad, and Rafael~A. Irizarry.
\newblock Frozen robust multiarray analysis (f{RMA}).
\newblock {\em Biostatistics (Oxford, England)}, 11(2):242--253, Apr 2010.

\bibitem{GSE104687}
Jeremy~A Miller, Angela Guillozet-Bongaarts, Laura~E Gibbons, Nadia Postupna,
  Anne Renz, Allison~E Beller, Susan~M Sunkin, Lydia Ng, Shannon~E Rose,
  Kimberly~A Smith, Aaron Szafer, Chris Barber, Darren Bertagnolli, Kristopher
  Bickley, Krissy Brouner, Shiella Caldejon, Mike Chapin, Mindy~L Chua,
  Natalie~M Coleman, Eiron Cudaback, Christine Cuhaciyan, Rachel~A Dalley, Nick
  Dee, Tsega Desta, Tim~A Dolbeare, Nadezhda~I Dotson, Michael Fisher, Nathalie
  Gaudreault, Garrett Gee, Terri~L Gilbert, Jeff Goldy, Fiona Griffin, Caroline
  Habel, Zeb Haradon, Nika Hejazinia, Leanne~L Hellstern, Steve Horvath, Kim
  Howard, Robert Howard, Justin Johal, Nikolas~L Jorstad, Samuel~R Josephsen,
  Chihchau~L Kuan, Florence Lai, Eric Lee, Felix Lee, Tracy Lemon, Xianwu Li,
  Desiree~A Marshall, Jose Melchor, Shubhabrata Mukherjee, Julie Nyhus, Julie
  Pendergraft, Lydia Potekhina, Elizabeth~Y Rha, Samantha Rice, David Rosen,
  Abharika Sapru, Aimee Schantz, Elaine Shen, Emily Sherfield, Shu Shi, Andy~J
  Sodt, Nivretta Thatra, Michael Tieu, Angela~M Wilson, Thomas~J Montine,
  Eric~B Larson, Amy Bernard, Paul~K Crane, Richard~G Ellenbogen, C~Dirk Keene,
  and Ed~Lein.
\newblock Neuropathological and transcriptomic characteristics of the aged
  brain.
\newblock {\em eLife}, 6:e31126, nov 2017.

\bibitem{GSE150961}
Angela Mo, Sini Nagpal, Kyle Gettler, Talin Haritunians, Mamta Giri, Yael
  Haberman, Rebekah Karns, Jarod Prince, Dalia Arafat, Nai-Yun Hsu, Ling-Shiang
  Chuang, Carmen Argmann, Andrew Kasarskis, Mayte Suarez-Farinas, Nathan
  Gotman, Emebet Mengesha, Suresh Venkateswaran, Paul~A. Rufo, Susan~S. Baker,
  Cary~G. Sauer, James Markowitz, Marian~D. Pfefferkorn, Joel~R. Rosh,
  Brendan~M. Boyle, David~R. Mack, Robert~N. Baldassano, Sapana Shah, Neal~S.
  LeLeiko, Melvin~B. Heyman, Anne~M. Griffiths, Ashish~S. Patel, Joshua~D. Noe,
  Sonia {Davis Thomas}, Bruce~J. Aronow, Thomas~D. Walters, Dermot~P.B.
  McGovern, Jeffrey~S. Hyams, Subra Kugathasan, Judy~H. Cho, Lee~A. Denson, and
  Greg Gibson.
\newblock Stratification of risk of progression to colectomy in ulcerative
  colitis via measured and predicted gene expression.
\newblock {\em The American Journal of Human Genetics}, 108(9):1765--1779,
  2021.

\bibitem{Panahi2021}
Bahman Panahi and Mohammad~Amin Hejazi.
\newblock Weighted gene co-expression network analysis of the salt-responsive
  transcriptomes reveals novel hub genes in green halophytic microalgae
  dunaliella salina.
\newblock {\em Scientific Reports}, 11(1):1607, Jan 2021.

\bibitem{Parsana2019}
Princy Parsana, Claire Ruberman, Andrew~E. Jaffe, Michael~C. Schatz, Alexis
  Battle, and Jeffrey~T. Leek.
\newblock Addressing confounding artifacts in reconstruction of gene
  co-expression networks.
\newblock {\em Genome Biology}, 20(1):94, 2019.

\bibitem{lr}
Chao-Ying~Joanne Peng, Kuk~Lida Lee, and Gary~M. Ingersoll.
\newblock An {I}ntroduction to {L}ogistic {R}egression {A}nalysis and
  {R}eporting.
\newblock {\em The Journal of Educational Research}, 96(1):3--14, 2002.

\bibitem{petal2015}
Juli Petereit, Sebastian Smith, Frederick~C. Harris, and Karen~A. Schlauch.
\newblock petal: {C}o-expression network modelling in {R}.
\newblock {\em BMC Systems Biology}, 10(2):51, 2016.

\bibitem{GSE107559}
Ralph~B. Puchalski, Nameeta Shah, Jeremy Miller, Rachel Dalley, Steve~R.
  Nomura, Jae-Guen Yoon, Kimberly~A. Smith, Michael Lankerovich, Darren
  Bertagnolli, Kris Bickley, Andrew~F. Boe, Krissy Brouner, Stephanie Butler,
  Shiella Caldejon, Mike Chapin, Suvro Datta, Nick Dee, Tsega Desta, Tim
  Dolbeare, Nadezhda Dotson, Amanda Ebbert, David Feng, Xu~Feng, Michael
  Fisher, Garrett Gee, Jeff Goldy, Lindsey Gourley, Benjamin~W. Gregor, Guangyu
  Gu, Nika Hejazinia, John Hohmann, Parvinder Hothi, Robert Howard, Kevin
  Joines, Ali Kriedberg, Leonard Kuan, Chris Lau, Felix Lee, Hwahyung Lee,
  Tracy Lemon, Fuhui Long, Naveed Mastan, Erika Mott, Chantal Murthy, Kiet Ngo,
  Eric Olson, Melissa Reding, Zack Riley, David Rosen, David Sandman, Nadiya
  Shapovalova, Clifford~R. Slaughterbeck, Andrew Sodt, Graham Stockdale, Aaron
  Szafer, Wayne Wakeman, Paul~E. Wohnoutka, Steven~J. White, Don Marsh,
  Robert~C. Rostomily, Lydia Ng, Chinh Dang, Allan Jones, Bart Keogh, Haley~R.
  Gittleman, Jill~S. Barnholtz-Sloan, Patrick~J. Cimino, Megha~S. Uppin,
  C.~Dirk Keene, Farrokh~R. Farrokhi, Justin~D. Lathia, Michael~E. Berens,
  Antonio Iavarone, Amy Bernard, Ed~Lein, John~W. Phillips, Steven~W. Rostad,
  Charles Cobbs, Michael~J. Hawrylycz, and Greg~D. Foltz.
\newblock An anatomic transcriptional atlas of human glioblastoma.
\newblock {\em Science}, 360(6389):660--663, 2018.

\bibitem{GSE115828}
Rinki Ratnapriya, Olukayode~A. Sosina, Margaret~R. Starostik, Madeline
  Kwicklis, Rebecca~J. Kapphahn, Lars~G. Fritsche, Ashley Walton, Marios
  Arvanitis, Linn Gieser, Alexandra Pietraszkiewicz, Sandra~R. Montezuma,
  Emily~Y. Chew, Alexis Battle, Gon{\c{c}}alo~R. Abecasis, Deborah~A.
  Ferrington, Nilanjan Chatterjee, and Anand Swaroop.
\newblock Retinal transcriptome and e{QTL} analyses identify genes associated
  with age-related macular degeneration.
\newblock {\em Nature genetics}, 51(4):606--610, Apr 2019.

\bibitem{GSE181225}
Sophie~E. Ruff, Nikita Vasilyev, Evgeny Nudler, Susan~K. Logan, and Michael~J.
  Garabedian.
\newblock P{IM}1 phosphorylation of the androgen receptor and 14-3-3 $\zeta$
  regulates gene transcription in prostate cancer.
\newblock {\em Communications Biology}, 4(1):1221, Oct 2021.

\bibitem{Russo2018}
Pedro S.~T. Russo, Gustavo~R. Ferreira, Lucas~E. Cardozo, Matheus~C. Bürger,
  Raul Arias-Carrasco, Sandra~R. Maruyama, Thiago D.~C. Hirata, Diógenes~S.
  Lima, Fernando~M. Passos, Kiyoshi~F. Fukutani, Melissa Lever, João~S. Silva,
  Vinicius Maracaja-Coutinho, and Helder~I. Nakaya.
\newblock {CEM}i{T}ool: a bioconductor package for performing comprehensive
  modular co-expression analyses.
\newblock {\em BMC Bioinformatics}, 19(1):56, 2018.

\bibitem{SChaefer2017}
Robert~J. Schaefer, Jean-Michel Michno, and Chad~L. Myers.
\newblock Unraveling gene function in agricultural species using gene
  co-expression networks.
\newblock {\em Biochimica et Biophysica Acta (BBA) - Gene Regulatory
  Mechanisms}, 1860(1):53--63, 2017.
\newblock Plant Gene Regulatory Mechanisms and Networks.

\bibitem{Song2012}
Lin Song, Peter Langfelder, and Steve Horvath.
\newblock Comparison of co-expression measures: mutual information,
  correlation, and model based indices.
\newblock {\em BMC Bioinformatics}, 13(1):328, Dec 2012.

\bibitem{GSE28435}
Kimberly~D Spradling, Lucille~A Lumley, Christopher~L Robison, James~L
  Meyerhoff, and 3rd Dillman, James~F.
\newblock Transcriptional analysis of rat piriform cortex following exposure to
  the organophosphonate anticholinesterase sarin and induction of seizures.
\newblock {\em Journal of neuroinflammation}, 8:83--83, 2011.

\bibitem{GSE44903}
Joachim Theilhaber, Sanjay~N. Rakhade, Judy Sudhalter, Nayantara Kothari, Peter
  Klein, Jack Pollard, and Frances~E. Jensen.
\newblock Gene {E}xpression {P}rofiling of a {H}ypoxic {S}eizure {M}odel of
  {E}pilepsy {S}uggests a {R}ole for m{TOR} and {W}nt {S}ignaling in
  {E}pileptogenesis.
\newblock {\em PLOS ONE}, 8(9):1--19, 09 2013.

\bibitem{Tieri2019}
Paolo Tieri, Lorenzo Farina, Manuela Petti, Laura Astolfi, Paola Paci, and
  Filippo Castiglione.
\newblock Network {I}nference and {R}econstruction in {B}ioinformatics.
\newblock In Shoba Ranganathan, Michael Gribskov, Kenta Nakai, and Christian
  Schönbach, editors, {\em Encyclopedia of Bioinformatics and Computational
  Biology}, pages 805 -- 813. Academic Press, Oxford, 2019.

\bibitem{Sipko2017}
Sipko van Dam, Urmo Võsa, Adriaan van~der Graaf, Lude Franke, and João~Pedro
  de~Magalhães.
\newblock Gene co-expression analysis for functional classification and
  gene–disease predictions.
\newblock {\em Briefings in Bioinformatics}, 19(4):575--592, 01 2017.

\bibitem{CoXpress}
M.~Watson.
\newblock Co{X}press: {D}ifferential co-expression in gene expression data.
\newblock {\em BMC Bioinformatics}, 7(509), 2006.

\bibitem{Zhang2005general}
Bin Zhang and Steve Horvath.
\newblock A general framework for weighted gene co-expression network analysis.
\newblock {\em Statistical applications in genetics and molecular biology},
  4(1), 2005.

\end{thebibliography}
\end{document}